\documentclass{emulateapj}
\include{mnras_jdf}
\usepackage[utf8]{inputenc}
\usepackage[english]{babel}
\usepackage{fontenc}

\usepackage{graphicx}
\usepackage{color}

\newcommand{\ZnII}{\hbox{{\rm Zn}{\sc \,ii}}}

\newcommand{\MnII}{\hbox{{\rm Mn}{\sc \,ii}}}
\newcommand{\MgI}{\hbox{{\rm Mg}{\sc \,i}}}
\newcommand{\MgII}{\hbox{{\rm Mg}{\sc \,ii}}}

\newcommand{\HI}{\hbox{{\rm H}{\sc \,i}}}

\newcommand{\Ha}{\hbox{{\rm H}$\alpha$}}

\newcommand{\flux}{erg\,s$^{-1}$\,cm$^{-2}$}

\newcommand{\mpy}{\hbox{M$_{\odot}$\,yr$^{-1}$}}

\newcommand{\msun}{\hbox{M$_{\odot}$}}

\newcommand{\ma}{\hbox{$\lambda 2796$}}
\newcommand{\mb}{\hbox{$\lambda 2803$}}
\newcommand{\mc}{\hbox{$\lambda 2852$}}

\newcommand{\zna}{\hbox{$\lambda 2026$}}
\newcommand{\mniia}{\hbox{$\lambda 2606$}}
\newcommand{\mniib}{\hbox{$\lambda 2594$}}
\newcommand{\mniic}{\hbox{$\lambda 2576$}}

\newcommand{\kms}{\hbox{${\rm km\,s}^{-1}$}}
\newcommand{\nn}{\nonumber}

\def\textbf#1{#1}

\begin{document}

\title{The VLT SINFONI \MgII\ Program for Line Emitters (SIMPLE) II: Background quasars probing $z\sim1$ galactic winds~\altaffilmark{1} }
\author{Ilane~{Schroetter\altaffilmark{2,3}},
        Nicolas~{Bouch\'e\altaffilmark{2,3}}, C\'eline P\'eroux\altaffilmark{4}, Michael T. Murphy\altaffilmark{5}, Thierry Contini\altaffilmark{2,3}, Hayley Finley\altaffilmark{2,3}}

\altaffiltext{1}{Based on observations made at the ESO telescopes under program 080.A-0364(A) 080.A-0364(B) and 079.A-0600(B).}
\altaffiltext{2}{CNRS/IRAP, 14 Avenue E. Belin, F-31400 Toulouse, France}
\altaffiltext{3}{University Paul Sabatier of Toulouse/ UPS-OMP/ IRAP, F-31400 Toulouse, France }
\altaffiltext{4}{Aix Marseille Universit\'e, CNRS, LAM (Laboratoire d’Astrophysique de Marseille) UMR 7326, 13388, Marseille, France.}
\altaffiltext{5}{Centre for Astrophysics and Supercomputing, Swinburne University of Technology, Hawthorn, Victoria 3122, Australia}


\begin{abstract}
The physical properties of galactic winds are of paramount importance for our understanding of galaxy formation.
Fortunately, they can be constrained using background quasars passing near star-forming galaxies (SFGs). 
From the 14 quasar--galaxy pairs in our VLT/SINFONI \MgII\ Program for Line Emitters (SIMPLE) sample, we reobserved the 10 brightest galaxies in \Ha\ with the VLT/SINFONI with 0.7" seeing and the corresponding
quasar with the VLT/UVES spectrograph. 
Applying geometrical arguments to these ten pairs, we find that 
four are likely probing galactic outflows, three are likely probing extended gaseous disks, 
and the remaining three are not classifiable because they are viewed face-on.
In this paper we present a detailed comparison between the line-of-sight kinematics and the host galaxy emission kinematics
 for the pairs suitable for wind studies.  
We find that the kinematic profile shapes (asymmetries) can be well reproduced by a purely geometrical  wind model with a constant wind speed, except for one pair (towards J2357$-$2736)
that has the smallest impact parameter $b=6$~kpc and requires an accelerated wind flow. 
Globally, the outflow speeds are $\sim$100~km~s$^{-1}$ and the mass ejection rates (or $\dot M_{\rm out}$) in the gas traced by 
the low-ionization species are similar to the star formation rate (SFR), meaning that the mass loading factor, $\eta =\dot M_{\rm out}/$SFR, is $\approx$1.0.
The outflow speeds are also smaller than the local escape velocity, which  implies that the outflows do not escape the galaxy halo and are likely to fall back into the interstellar medium.
\end{abstract}


\section{Introduction}

  \textbf{
Currently, under the Cold Dark Matter (CDM)  scenario, galaxies form via 
the growth of initial density fluctuations. This scenario is very successful
because the observed large scale structure is well matched
by the clustering of halos in N-body simulations \citep[e.g.][]{springel_06}.
With the help of state-of-the art hydrodynamical simulations  \citep{genel_14,muratov_15,schaye_15},
this scenario has recently 
successfully reproduced more complex observables
such as the galaxy morphologies \citep{genel_14, Vogelsberger_14a}. 
}

 \textbf{
A major success of these recent hydro-simulations is a better understanding of disk formation  at high-redshifts $z>1$. 
Indeed, these simulations \citep{genel_12,genel_14} are in good agreement with the morphologies and kinematic observed for Lyman break galaxies, which appear to be dominated by
 gas-rich rotating disks, based on data 
from deep HST observations \citep{elmegreen_07,wuyts_11} and large  Integral Field Units (IFU) surveys like SINS
\citep{forsterschreiber_06,genzel_08, forsterschreiber_09} and MASSIV \citep{epinat_12,contini_12}.
}

 \textbf{
One major problem remains, however, namely that  the luminosity function for low mass galaxies ($L<L_*$) is difficult to reproduce.
For instance, the Illustris simulations \citep{genel_14} over-predicts the number of $z=0$ low-mass galaxies despite implementing strong galactic outflows \citep[but see][]{schaye_15}. This tension can be recast in terms of the `galaxy formation efficiency'  
\citep[][]{moster_10, papastergis_12, leauthaud_11, leauthaud_12, behroozi_10},
 which is maximal for $\sim L_*$ galaxies and steeply decreases in the low-mass regime ($L<L_*$).
In the low-mass regime, galactic winds, created by accumulated supernova explosions, are commonly invoked to 
expel baryons back into the inter-galactic medium \citep[][]{white_91, dekel_86}  since baryons in these galaxy halos are expected to cool rapidly  \citep{white_78, white_91,keres_05, dekel_09}.
}

\textbf{Although} galactic winds seem to occur in every star forming galaxy,  
their properties remain poorly 
 constrained despite many attempts at characterizing them \citep[][]{heckman_96, heckman_00, martin_98, martin_99, rupke_05, rubin_10, martin_12}. This lack of knowledge 
prevents us from correctly modeling galaxies in numerical simulations, which often require ad-hoc recipes
 \citep[][]{oppenheimer_06, oppenheimer_10, dubois_08, roskar_13, rosdahl_13, shen_12, shen_13, dekel_13}.
In particular, the best estimates for the ejected mass rate ($\dot M_{\rm out}$) \textbf{using standard galaxy absorption lines} are uncertain by orders of magnitude 
\citep[e.g.][]{heckman_90, heckman_00, pettini_02, martin_02, martin_05, martin_12, martin_13}.

The main reason for the large uncertainties is that traditional spectroscopy does not give information of the material physical location because the gas could 
be at a distance of 100~pc, 1~kpc, or 10~kpc from the galaxy.
Indeed, the standard method usually uses the galaxy spectrum \textbf{and in some cases} stacked galaxy spectra, to obtain the absorption lines corresponding to the outflowing materials.
However, background quasars have been recently used to constrain the properties of winds \citep[][]{bouche_12, kacprzak_14} using  low-ionization 
absorption lines, like \MgII\ ($\lambda 2796, \lambda 2803$).
Indeed, when the quasar apparent location is close to the galaxy minor axis,  the line of sight (LOS) \textbf{is expected} to intercept the wind. 
Thus background quasars give  us the three main ingredients necessary for determining accurate ejected mass rates: the gas localization (impact parameter), the gas 
column density and the wind radial  (de-projected) velocity, provided that the galaxy inclination is known. 

The background quasar technique also provides the ability to  better constrain the ejected mass outflow rate and its relation to the galaxy star formation rate (SFR) 
via the so-called mass loading factor $\eta\equiv\dot M_{\rm out}/$SFR, which is a critical ingredient  for numerical simulations \citep{oppenheimer_06, oppenheimer_10, dekel_13}.
\textbf{As opposed to relying only on the galaxy spectra to study outflows, }
the background quasar method has several advantages:
it gives us a more precise location of the absorbing gas relative to the galaxy and, because the quasar is seen as a point source, it also provides us with a good characterization of the point spread function (PSF), 
an important ingredient \textbf{for} deriving the intrinsic galaxy properties \textbf{from} integral field unit (IFU) data.

Recently, there has been 
progress in this field with low-redshift $z\sim0.1$ star-forming galaxies \citep[][]{bouche_12, kacprzak_14} applying this technique.
In this paper, we use the sample of 14 intermediate-redshift $z\sim1$ galaxy-quasar pairs from the SINFONI \MgII\ Program for Line Emitters  \citep[SIMPLE, ][hereafter paper~I]{bouche_07}
to constrain the outflow properties (e.g. mass ejection rate, outflow velocity) of star-forming galaxies when the quasar is suitably located relative to the foreground galaxy.

The outline of this paper is as follows:
\S~2 describes the sample and the new VLT SINFONI/UVES data acquired.
\textbf{In \S~3, we present the analysis of the SINFONI and UVES data together with the selection of pairs suitable for wind studies (wind-pairs).}
\textbf{In \S~4, we describe our wind model and the derived outflow rates for the wind-pairs.}
We end with our conclusions and discussions in \S~5.
In this study we used the following cosmological parameters: $H_0$=70~km~s$^{-1}$, $\Omega_{\Lambda}$=0.7 and $\Omega_{M}$=0.3.


\section{The SIMPLE sample}

Because the probability of finding galaxy-quasar pairs is very low, one must \textbf{employ targeted} strategies \textbf{for gathering a} suitable sample of galaxy-quasar pairs to study the properties
of circumgalactic gas around galaxies, which can lead to constraints on outflows \citep[][]{bouche_12, kacprzak_14} or inflows \cite[][]{bouche_13, peroux_13}.
We thus designed the SIMPLE survey to build a sample of intermediate redshift $z\sim1$ quasar-galaxy pairs  (paper~I).

The SIMPLE sample \citep{bouche_07} was selected with the following criteria: 
the rest-frame equivalent width of \textbf{intervening \MgII\ ($W^{\lambda2796}_{r}$) absorptions detected in background quasar spectra} had to be at least 2 \AA. 
This criterion ensures that  the \textbf{associated galaxies} will be at small impact parameters ($b<3$"), 
given the $W_r$--impact parameter anti-correlation  \citep[e.g.][]{steidel_95, bouche_06, menard_09, chenhw_12}, \textbf{and thus that they} will be located within 
the field-of-view of the IFU SINFONI (8" each side).
Moreover, the absorber's redshift must be $0.8 < z < 1.0$ so that the \Ha\ emission line falls inside the SINFONI J band.
\textbf{These criteria led to the detection of 14 galaxies out of 21 (70\%\ success rate)  \citep{bouche_07}.}

The SINFONI data presented in \citet{bouche_07} were shallow with exposure times $\leq~40$ min and seeing conditions $>0.8$".
Since we aim to precisely compare the host galaxy kinematics (derived from the \Ha\ emission line) with  the kinematics of the absorbing material measured in the quasar line-of-sight, 
we acquired new   VLT/SINFONI and VLT/UVES data.
From the sample of 14 galaxies in \citet{bouche_07}, we re-observed a sub-sample of 10 galaxies, those with the highest initial \Ha\ fluxes, with longer integration times (2-3 hr) and in better seeing conditions ($<0.8"$). 

\textbf{The SINFONI observations, done in service mode, }
were  optimized  by adopting a  `on source'
dithering strategy designed to ensure a continuous integration at the host location. 
The UVES \citep{dekker_00} data were taken in both visitor mode and service mode.


\subsection{SINFONI Data Reduction}

The data reduction was performed  as in \citet{bouche_07,ForsterSchreiberN_09a,bouche_12},
 using the  SINFONI pipeline \citep[SPRED,][]{SchreiberJ_04a,AbuterR_06a} complemented with  custom routines
such as
the OH sky line removal scheme of \citet{DaviesR_06a} and the Laplacian edge cosmic ray removal technique of \citet{vanDokkumP_01a}.

Regarding the wavelength calibration, we emphasize that we applied the heliocentric correction to the sky-subtracted frames, and 
each frame was \textbf{associated with} a single reference frame  by cross-correlating each of the science frames spectrally against the reference frame
(the first science exposure).   
For each observing block, we  use the the quasar continuum to spatially register the various sets of observations.  
Finally, \textbf{we created a} co-added cube from all the individual
sky-subtracted 600s exposures using a median clipping at 2.5$\sigma$.

\textbf{Flux calibration} was performed on a night-by-night basis using the
broadband magnitudes of the standards from 2MASS. The flux calibration is accurate to
$\sim15$\%.  
Finally, the atmospheric transmission was calibrated out by dividing the science
cubes by the integrated spectrum of the telluric standard.
 
 \textbf{In Figures~\ref{fig:flux_1} and~\ref{fig:flux_4} we present the flux, velocity and dispersion 
maps for each galaxy.}

\begin{table*}
\centering
\caption{Summary of SINFONI 080.A-0364(B) observations. \label{table:obs}}
\begin{tabular}{lllccc}
\hline
Field	 		          & $z_{\rm qso}$    & $W_r^{\lambda2796}$(\AA) 	  &  PSF($\arcsec$)   &  T$_{\rm exp}$(s) &  Date     \\
(1)   		                  &	(2)	     & (3)	           &  (4)             &  (5)       &  (6)	       \\
\hline
J0147+1258                        &1.503             &4.025                &0.6              &9600                & 2007-10-12 2008-01-03,04,09   \\
J0226-2857                        &2.171             &4.515                &0.6              &9600                & 2007-10-06 2008-01-03,05   \\
J0302-3216                        &0.898             &2.27                 &0.7              &7200                & 2007-10-02   \\
J0448+0950                        &2.115             &3.169                &0.8              &4800                & 2007-12-04,16    \\
J0822+2243                        &1.620             &2.749                &0.8              &4800                & 2007-12-18 2008-01-03   \\
J0839+1112                        &2.696             &2.316                &0.8              &4800                & 2007-12-14,15,23 2008-01-01   \\
J0943+1034                        &1.239             &3.525                &0.6              &7200                & 2007-12-22 2008-01-06   \\
J1422-0001                        &1.083             &3.185                &0.7              &9600                & 2008-02-15,25 2008-03-14   \\
J1441+0443                        &1.112             &2.223                &0.6              &12000               & 2008-03-14,15,25   \\
J2357-2736                        &1.732             &1.940                &0.6              &7200                & 2007-10-02    \\
\hline
\end{tabular}\\
\vspace*{0,1cm}
{
(1) Quasar name;
(2) Quasar emission redshift;
(3) \MgII\ rest-equivalent width; 
(4) FWHM of the seeing PSF;
(5) Exposure time;
(6) Dates of observations.
}\\   
\end{table*}


\subsection{UVES Observations}

The UVES data were taken during two distinct observing runs: 13 hrs in Service Mode (ESO 79.A-0600) and 1.5n in Visitor Mode (ESO 80.A-0364). 
We used a combination of 390 + 564, 390 + 580 and 390 + 600 nm central wavelength settings appropriate to the range of wavelengths for the lines we
were seeking.  The total exposure time for each object was split
into two or three equal observing blocks to minimize the effect of
cosmic rays.  The slit width was 1.2 arcsec, yielding a spectral resolution $R=\lambda$/$\Delta \lambda$ $\sim$ 45 000. 
A 2 $\times$ 2 CCD binning was used \textbf{for all observations}. 
The observational set-ups are
summarized in Table~\ref{table:obsuves}.  

The data were reduced using version 3.4.5 of the UVES pipeline in MIDAS.  Master bias and flat images were constructed using
calibration frames taken closest in time to the science frames.  The
science frames were extracted with the optimal option.  The blue \textbf{portion} of the spectra was checked order by order to verify that all were properly extracted. The
spectra were then corrected to the vacuum heliocentric reference frame. 
The resulting spectra were combined, weighting each spectrum with its signal-to-noise ratio. 
To perform 
\textbf{absorption line analysis}, the spectra were
normalized using cubic spline functions of the orders of 1--5 as the local
continuum. In this paper, we present the UVES data for the four pairs that will be classified as \textbf{pairs suitable for wind studies, hereafter wind-pair}. 
The remainder will be presented in subsequent papers.
\textbf{We find important to mention that UVES and SINFONI data have their wavelength calibrations made in vacuum.}

\begin{table*}
\centering
\caption{Summary of UVES observations. \label{table:obsuves}}

\begin{tabular}{l c c c c  }
\hline
Target 		&setting $\lambda_c$ (nm)	&  T$_{\rm exp}$ (s)	&Run ID$^a$	&Date 		\\
\hline
J0147+1258	&390+580		&4440  &SM/079.A-0600(B)			&2007-07-23  2007-08-14	\\
J0226-2857	&390+580		&9000  &SM/079.A-0600(B)			&2007-07-24,27  2007-09-04	\\
J0302-3216	&390+564		&5430  &SM/079.A-0600(B)			&2007-08-02		\\
J0448+0950	&390+564		&13200 &VM/080.A-0364(A)			&2008-01-28,29		\\
J0822+2243	&390+564		&7200  &VM/080.A-0364(A)			&2008-01-29		\\
J0839+1112	&390+564		&13200 &VM/080.A-0364(A)			&2008-01-28		\\
J0943+1034	&390+580		&9000  &SM/079.A-0600(B)			&2007-04-18,22		\\
J1422-0001	&390+564		&9000  &SM/079.A-0600(B)			&2007-04-12,14		\\
J1441+0443	&390+600		&8100  &VM/080.A-0364(A)			&2008-01-28,29		\\
J2357-2736	&390+564		&4440  &SM/079.A-0600(B)			&2007-05-15		\\
\hline
\end{tabular}

\vspace*{0,1cm}
$^a$ SM stands for Service Mode and VM for Visitor Mode.\\
\end{table*}


\subsection{Ancillary data}
\textbf{For all of the galaxy$-$quasar pairs, we checked for ancillary data and found two pairs with available HST observations imaging. The first one is J0448+0950 which has {\it HST}/WFPC2 (F555W filter) data from \citet[][HST proposal ID 5393]{lehnert_99}.  The second one is J0839+1112 which has {\it HST}/WFPC2 
(F702W filter) from {\it HST} proposal ID 6557 (PI: Steidel) first published in \citep{Kacprzak_10a}.
These HST data are discussed later and shown in Figure~\ref{fig:flux_1}.}


\section{Results}


\subsection{Galaxy Emission Kinematics}
\label{section:kinematics}

In most cases, the PSF cannot be \textbf{estimated} from the data itself given the small SINFONI IFU field-of-view (8x8 arcsec$^2$).
Here, one advantage of using galaxy-quasar pairs is that the knowledge of the PSF \textbf{can be determined from} the quasar continuum in the data cube.
This information is \textbf{crucial for deriving} intrinsic values of host galaxy parameters.
Moreover, fitting a disk model to seeing-limited data requires good knowledge of the PSF (see \citet{cresci_09} and \citet{epinat_12}).

From IFU data, it is customary to extract moment maps  (e.g. flux, velocity  and dispersion maps)  from the emission line(s) spectra.
This is usually done on a pixel by pixel basis, as most algorithms treat the spaxels to be independent \citep[e.g.][]{forsterschreiber_09, cresci_09, epinat_09, law_07, law_09},
a condition that requires high quality data with a high signal-to-noise ratio in each spaxel, in order to constrain the width and centroid of the emission lines.
Here, we \textbf{avoid shortcomings of the traditional techniques by comparing} the three-dimensional data cubes directly to a three-dimensional galaxy disk model 
using the  Galpak$^{\rm 3D}$ tool  \citep{bouche_15}.  
The algorithm \textbf{models} the galaxy directly in 3D (x, y, $\lambda$), \textbf{and the model} is then convolved  with the atmospheric PSF and the instrumental line spread function [LSF]. 
The (intrinsic) model parameters are optimized using Monte Carlo Markov Chains (MCMC), from which we compute the posterior distributions on each of the parameters. 
The form of the rotation curve $v(r)$ is given by $v(r)=V_{\rm max}\;2/\pi\; \arctan(r/r_t)$,
 where $r_t$ is the turn-over radius and $V_{\rm max}$ the maximum rotation velocity.
The algorithm has several advantages: (i) the dynamical center does not need to be fixed spatially, and (ii) the SNR required per spaxel for the creation of 2D velocity maps is relaxed.
In addition, since the actual PSF is well known from the quasar continuum, the returned parameters, \textbf{including} the galaxy position angle (PA\textbf{, which is defined by the angle between the celestial 
North and the galaxy major axis, anticlockwise}), 
inclination (i), size, maximum rotation velocity ($V_{\rm max}$)~\footnote{Since the three-dimensional disk model is inclined, 
the value $V_{\rm max}$ is the deprojected maximum velocity, corrected for inclination.}, are intrinsic (or deconvolved) galaxy parameters.
Extensive tests presented in Bouch\'e et al. (2014, submitted) show that the algorithm requires data with a SNR$_{\rm max}>3$ in the brightest pixel. 
For high SNR, all parameters can be well recovered, but in low SNR data, degeneracies can appear: \textbf{for instance between} turn-over radius and $V_{\rm max}$. 

In order to first assess the flux profile properties, exponential vs. gaussian surface brightness profile and axis-ratios, we analyzed the collapsed cubes (i.e. line integrated, continuum-subtracted) 
with the galfit2D tool. This tool is our custom 3D equivalent to Galfit from \citep{peng_10}, which fits isophotes to the images (with the PSF convolution) 
\textbf{and then uses these isophotes to compute} the radial surface brightness profile.
\textbf{With the results from the 2D algorithm, we obtain an initial indication of the galaxy inclination}
 from the axis-ratio and the profile shape (exponential vs. gaussian), before analyzing the kinematics in the 3D data.
We find that seven \textbf{galaxy surface-brightness profiles} can be described by Exponential profiles while three are best described by a Gaussian profile (e.g J1422-0001).

Using the GalPaK$^{3D}$ tool, we fit the kinematics directly to the data-cubes. The results are shown in Table~\ref{table:galpak}.
We emphasize that the surface-brightness profile breaks the common inclination-$V_{\rm max}$ degeneracy in kinematic analysis.
For every galaxy, we checked that the MCMC chain converged for each of the parameters and \textbf{estimated the uncertainties from the last 60 percents of the iterations. }
For J1422$-$0001, some of the kinematic parameters remain unconstrained, 
because the rotation curve appears shallow such that the turn-over radius $r_t$ and the circular velocity $V_{\rm max}$ are degenerate.
The parameters relevant for this study \textbf{for defining the kinematic major axis} (PA) are well constrained, however.

As we will see in section 3.4 (also illustrated in Figure~\ref{fig:models}),  galaxy inclination is a critical parameter for the wind model 
(details on the model are described in \S~\ref{section:wind}).
We cross-check the inclination measured using various methods (mainly galfit2D and GalPaK$^{3D}$) and from the SINFONI and HST data sets when present.
In particular, for J0943+1034, the galaxy's inclination is set to the value obtained from the 2D profile fitting since GalPaK$^{3D}$ did not converged for the turn-over radius parameter. 
For J0839+1112, the galaxy's inclination is set to the value obtained from the archival HST image. 
For J0448+0950, the  galaxy's inclination obtained from the SINFONI and HST data are consistent. 

In \textbf{Figures~\ref{fig:flux_1}  we present the data, ie. the observed flux and kinematics maps, and the fitted model. 
For each galaxy, the SINFONI data are shown in the  first row, along with the HST image when available.
In the second row, we present the results from the 3D kinematic fitting with the GalPaK$^{3D}$ algorithm,
where we show the dispersion, velocity, flux and residuals maps, from right to left. }
\textbf{The residuals maps are generated from the residuals cubes which are just the difference between the data and the model normalized by the pixel noise.
The  2D maps show  the mean of the residuals in each spaxel normalized by its standard error. 
All the panels have the north up and the east to the left.}

\begin{table*}
\centering
\caption{Kinematic and morphological parameters.
\label{table:galpak}}
\begin{tabular}{lcccccccccc}
\hline
Galaxy    & b (kpc)  &$\alpha$ ( $\mathring{}$ )  & inclination ( $\mathring{}$ )        & PA ( $\mathring{}$ )      & flux                   & $V_{\rm max}$        & redshift & r$_{1 \over 2}$ & Profile & Class \\
(1) & (2) & (3) & (4) & (5) & (6) & (7) & (8) & (9) & (10) & (11)\\
\hline
J0147+1258     &17.9$\pm$1.02      &30$\pm$30	                          & 24.4$\pm$3.3                         & $-$69$\pm$3                  & $1.63\cdot10^{-16}$    & 241$\pm$38        &1.0389  & 7.11 $\pm$ 0.20  & EXP &face-on \\
J0226-2857     &$\le$ 2.0$\pm$1.01 &56$\pm$3.0                          & 47.9$\pm$1.0                         & 91$\pm$1                  & $2.01\cdot10^{-16}$    & 50$\pm$12         &1.0223 & 2.69 $\pm$ 0.04   & EXP &Ambig. \\
J0302-3216     &19.7$\pm$0.95      &16$\pm$15                          & 30.4$\pm$1.5                         & $-$37$\pm$3                  & $2.70\cdot10^{-16}$    & 180$\pm$15        &0.8223  & 8.99 $\pm$ 0.31  & EXP &face-on \\
J0448+0950     &13.7$\pm$0.96      &79$\pm$3.0                          & 52.0$\pm$1.2                         & 31$\pm$1                 & $5.03\cdot10^{-16}$    & 253$\pm$10         &0.8391 & 7.85 $\pm$ 0.07   & EXP &wind-pair \\
J0822+2243     &21.8$\pm$0.95      &32$\pm$30                          & 17.9$\pm$0.7                         & 168$\pm$1                & $5.04\cdot10^{-16}$    & 328$\pm$14        &0.8102  & 4.14 $\pm$ 0.06  & EXP &face-on \\
J0839+1112     &26.8$\pm$0.94      &59$\pm$6.0                          & 72$\pm$5$^\dagger$                       & 139$\pm$4                & $1.53\cdot10^{-16}$    & 113$\pm$20        &0.7866 & 5.65 $\pm$ 0.29    & EXP &wind-pair \\
J0943+1034     &24.3$\pm$1.01      &32$\pm$3.0                          & 43$\pm$5$^{\dagger\dagger}$                           & 140$\pm$1                 & $3.81\cdot10^{-16}$    & 327$\pm$10        &0.9956   & 8.73 $\pm$ 0.21 & EXP &inflow-pair \\
J1422-0001     &12.7$\pm$0.98      &17$\pm$5.0                          & 55$\pm$5                           & 81$\pm$3                 & $8.93\cdot10^{-17}$    & 130$\pm$20$^{\dagger\dagger\dagger}$        &0.9096   & 4.30 $\pm$ 0.16 & GAU &inflow-pair \\
J1441+0443$^{\dagger\dagger}$      &10.1$\pm$1.02      &90$\pm$6.0                          & $\cdots$                         & 87$\pm$4                 & $6.62\cdot10^{-17}$    & $\cdots$         &1.0384  & 2.99 $\pm$ 0.18 & GAU &wind-pair\\
J2357-2736     &6.7$\pm$0.95       &68$\pm$4.0                          & 51.6$\pm$2.2                         & 109$\pm$2                & $1.29\cdot10^{-16}$    & 187$\pm$15        &0.8149  & 5.53 $\pm$ 0.14  & GAU &wind-pair \\
\hline
\end{tabular}\\
\vspace*{0,1cm}
{
(1) Quasar name;
(2) Impact parameter;
(3) Azimuthal angle $\alpha$ (Section~\ref{section:azimuthal});
(4) Galaxy inclination (degrees);
(5) Position Angle (degrees);
(6) Integrated \Ha\ flux of the galaxy (\flux);
(7) Maximum rotation velocity (\kms);
(8) H$_{\alpha}$ redshift (see \S~\ref{section:redshifts});
(9) Half-light radius (kpc);
(10) Assumed flux profile (Exp. or Gau.);
(11) Class (inflow-pair/wind-pair) based on $\alpha$ selection.
$^\dagger$ the inclination is determined from the HST  data.
$^{\dagger\dagger}$ galaxy parameters are derived from 2D fitting (galfit2D).
$^{\dagger\dagger\dagger}$    turn-over radius is fixed to $r_t/r_{1/2}=0.25$.
}
\end{table*}


\begin{figure*}[h!]
  \centering
  \includegraphics[width=19cm]{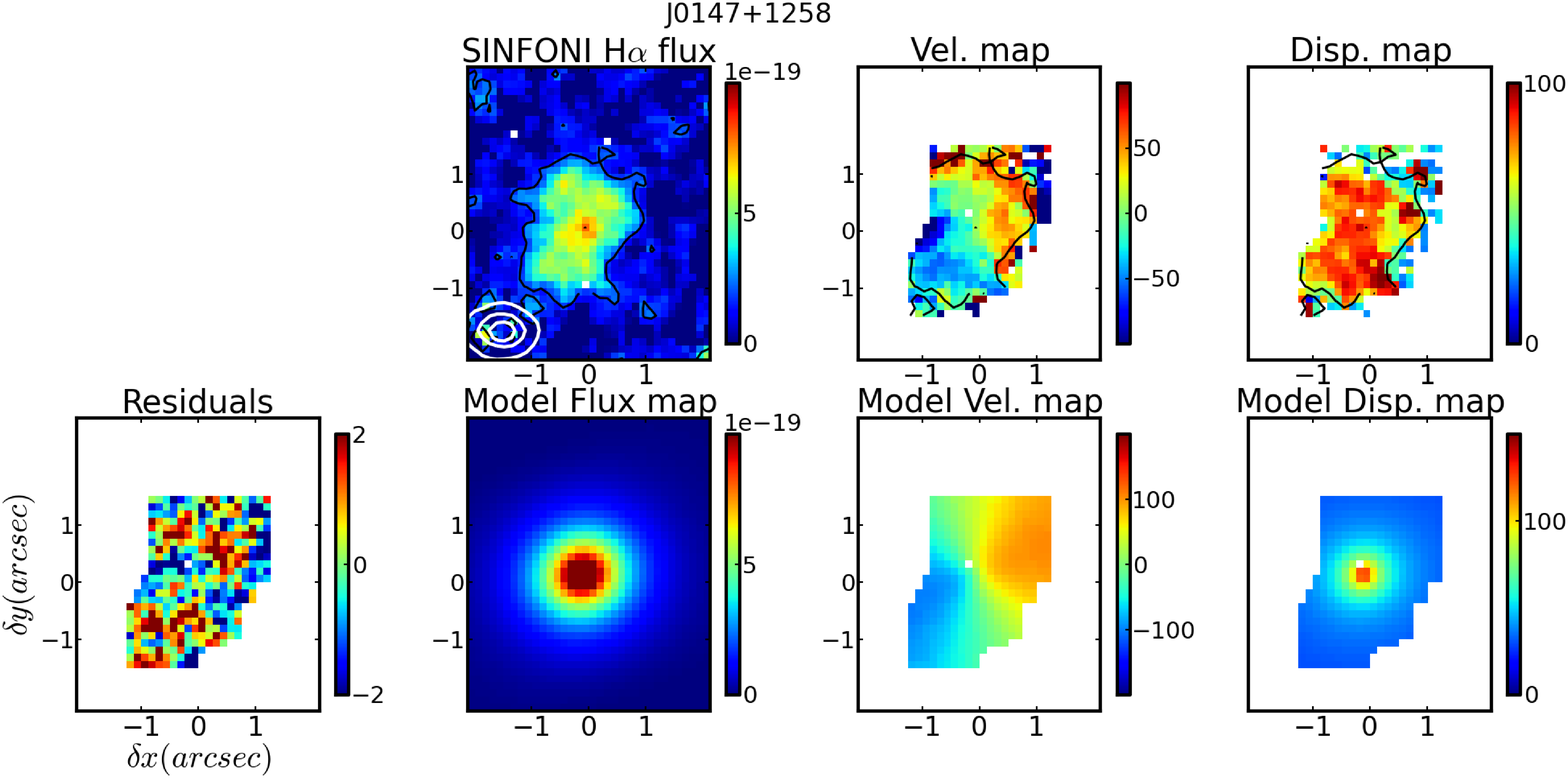}
  \includegraphics[width=19cm]{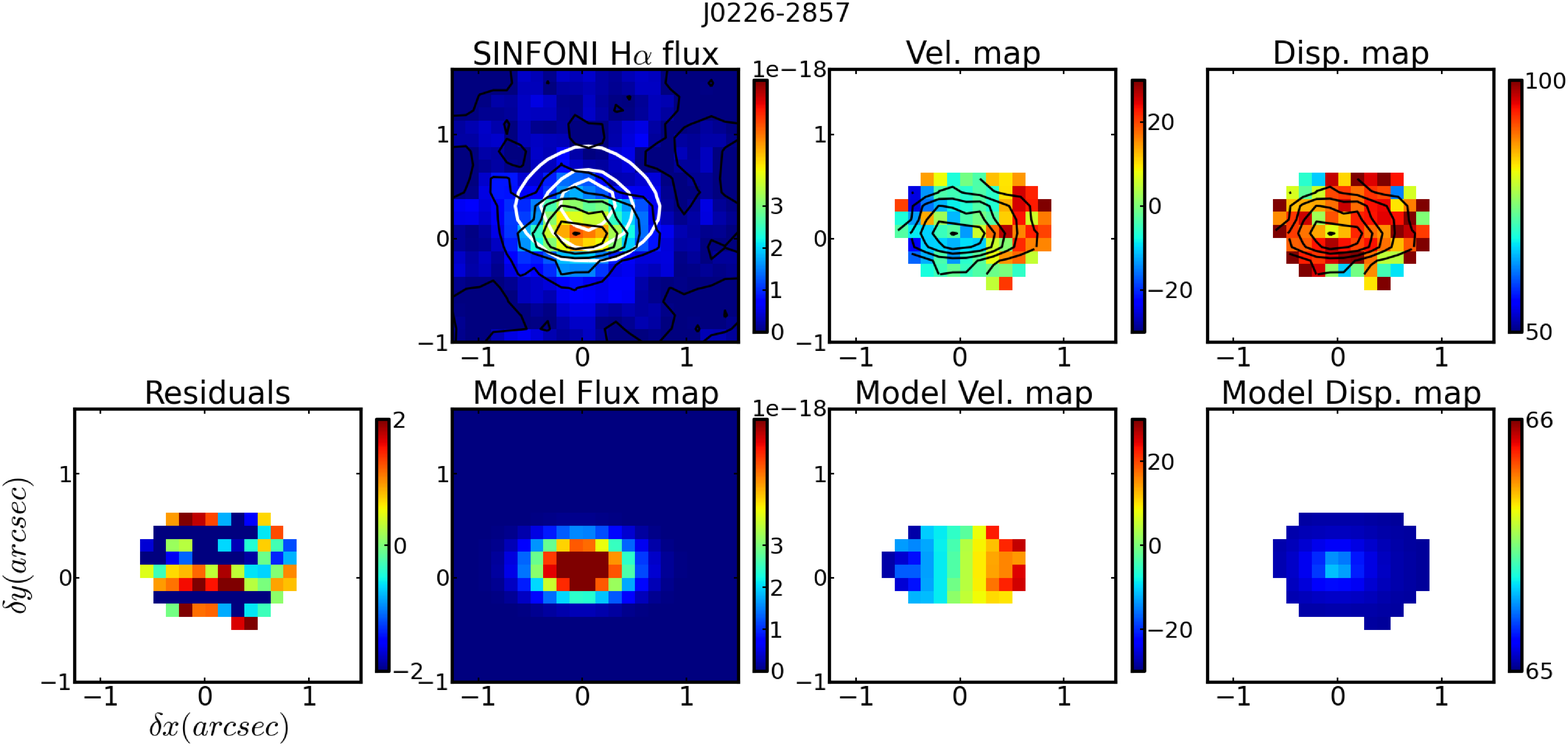}
  \includegraphics[width=19cm]{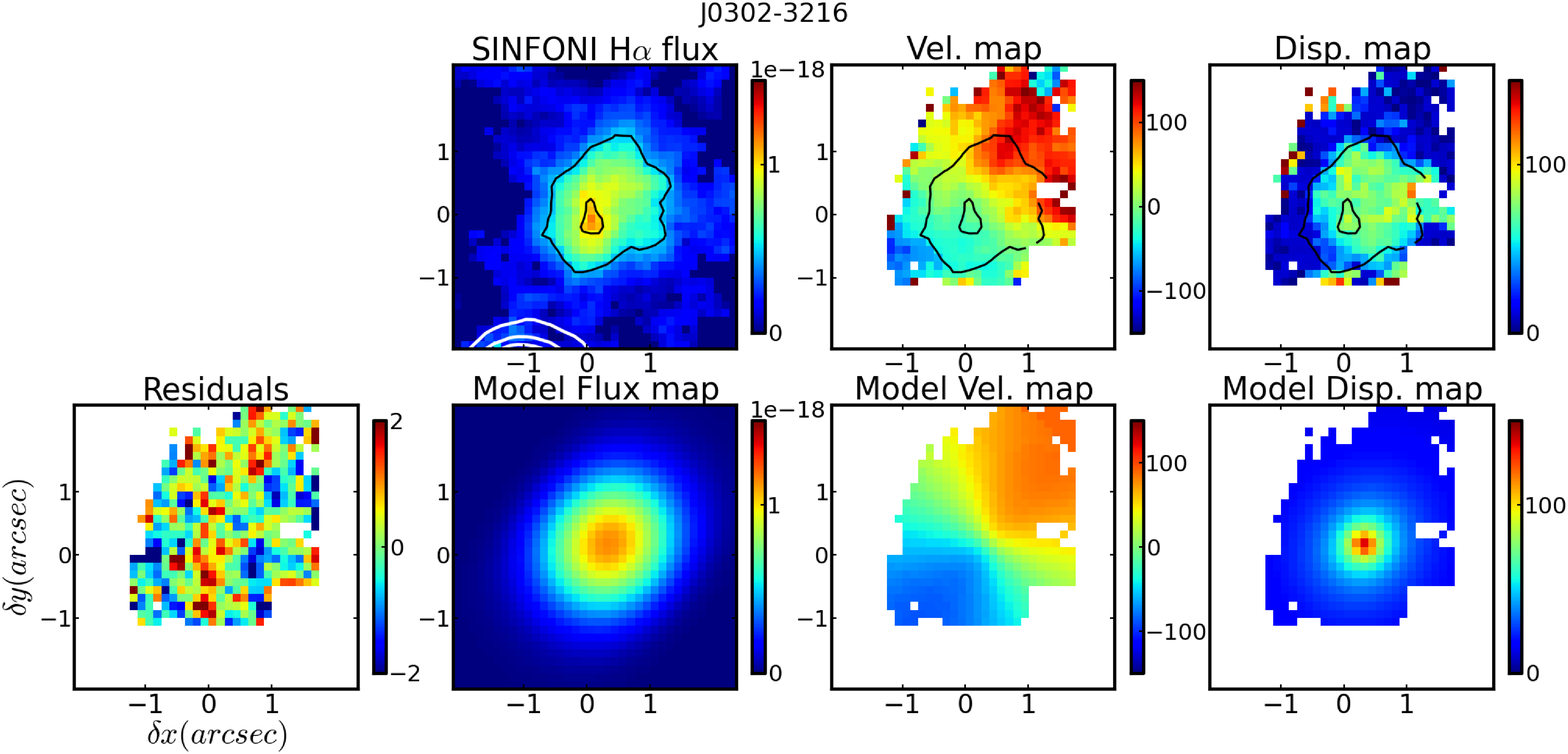}
  \caption{\textit{From left to right:} top: The HST/WFPC2 image (when available), the \Ha\ J-band SINFONI flux (\flux), the velocity map (in \kms) derived from the SINFONI data and the dispersion map (in \kms).
  bottom: \textbf{the residuals cube represented in 2D} (in $\sigma$), the intrinsic reconstructed galaxy with GalPaK$^{3D}$ (deconvolved from the PSF given by the quasar), its velocity map and the dispersion map.
  The quasar position is represented by the white contours on the observed flux maps when present in the map.
  \textbf{In each panel, north is up and east is to the left.}
  }
  \label{fig:flux_1}
\end{figure*}
\addtocounter{figure}{-1}
\begin{figure*}[]
  \centering
  \includegraphics[width=19cm]{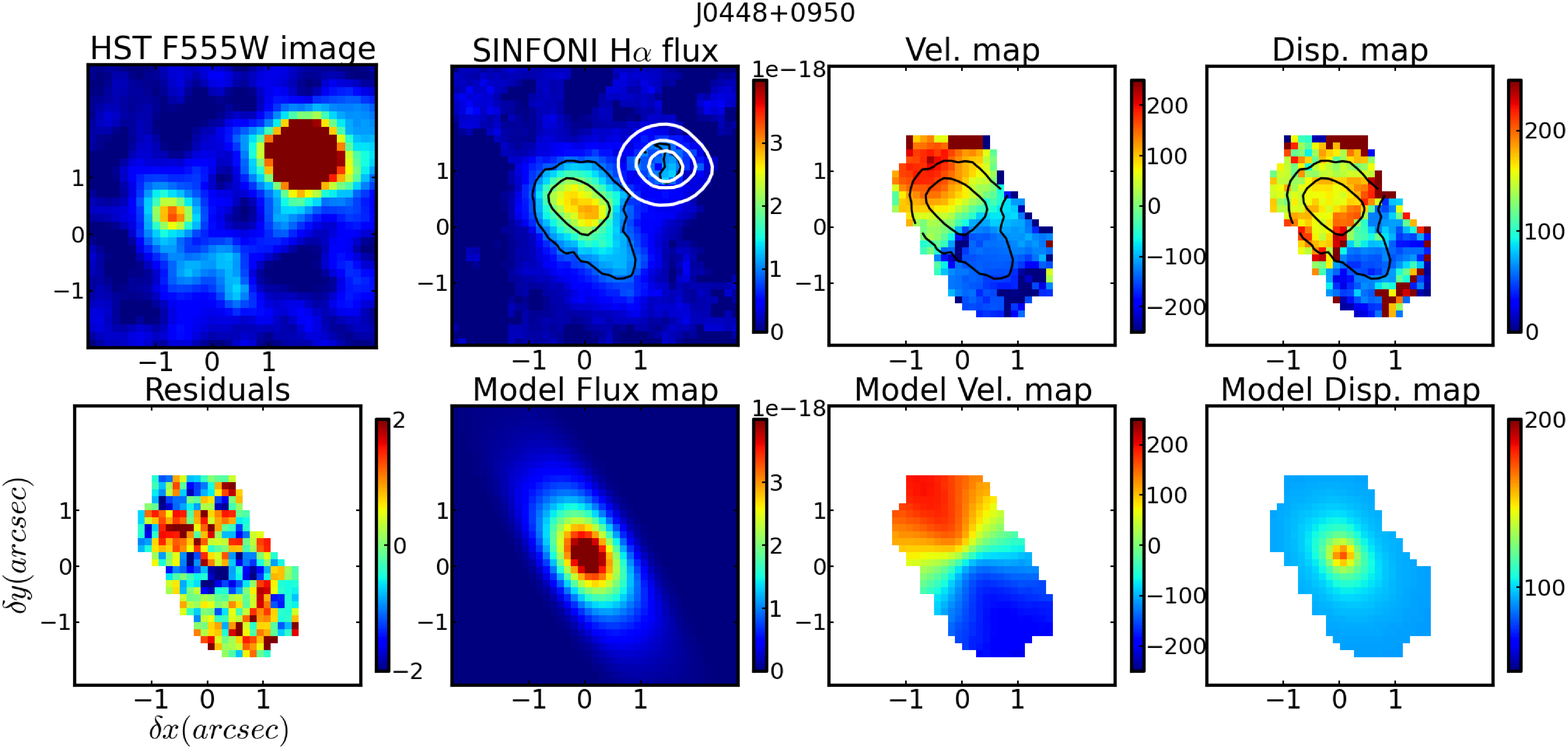}
  \includegraphics[width=19cm]{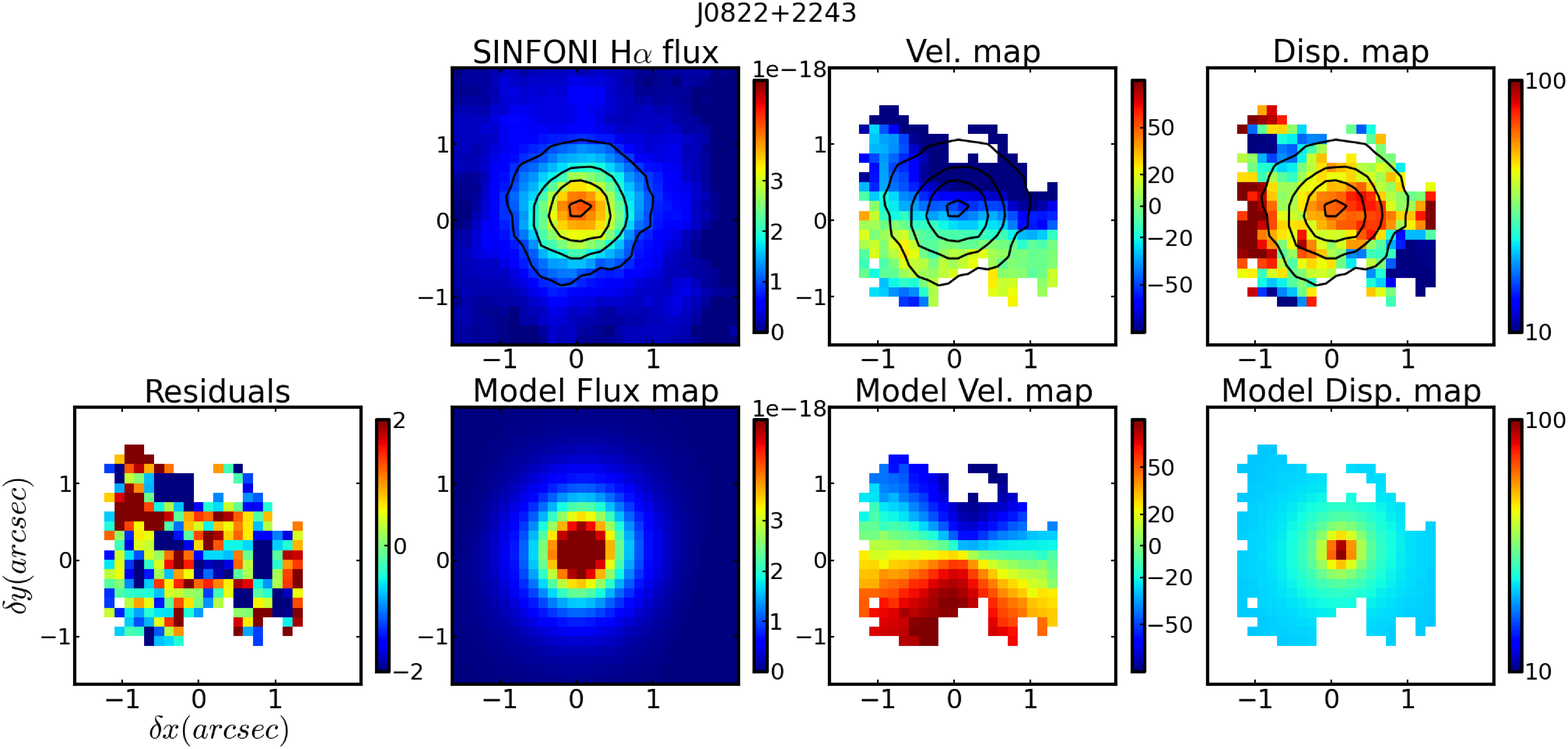}
  \includegraphics[width=19cm]{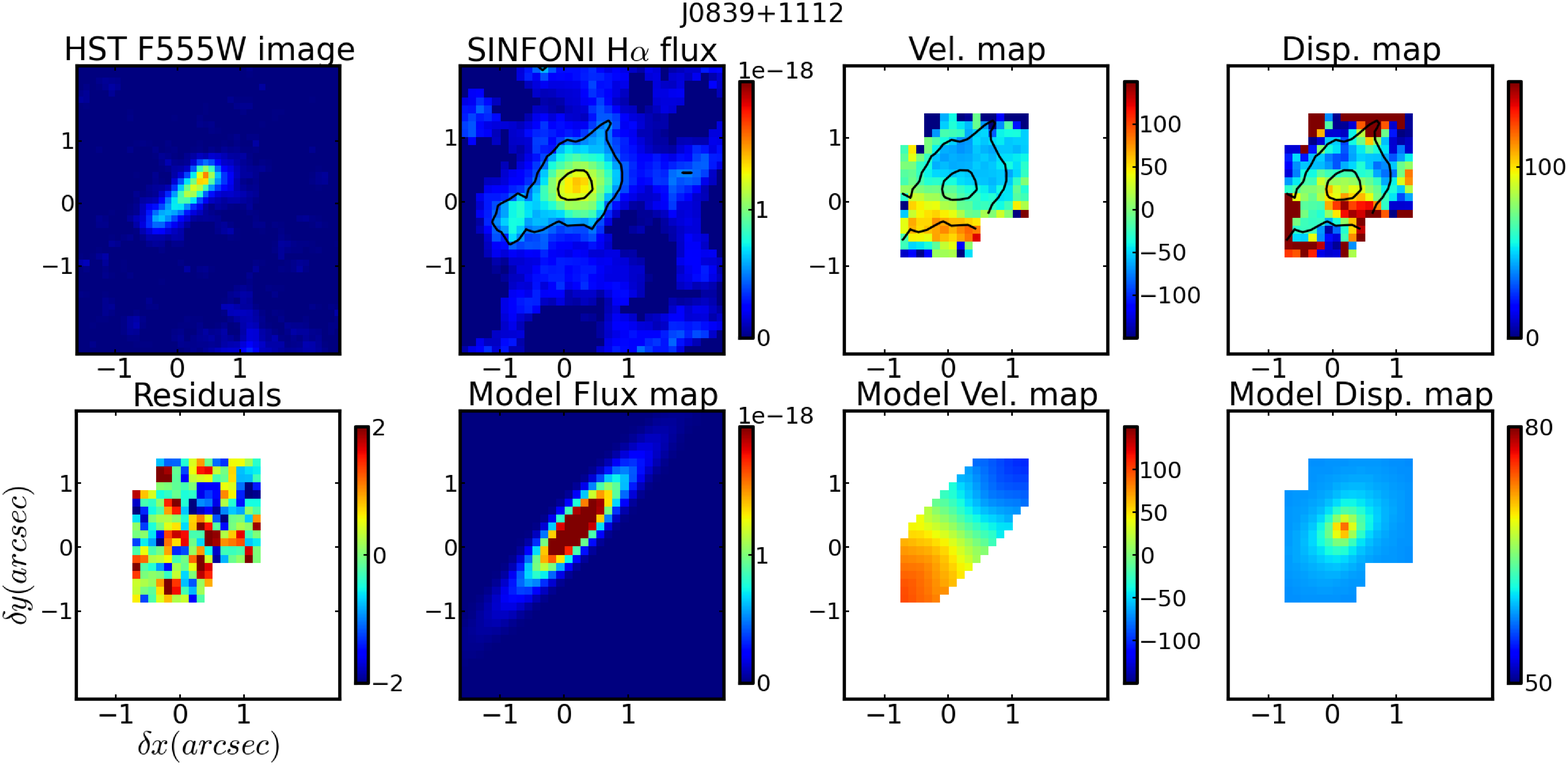}
  \caption{(continued)  }
  \label{fig:flux_2}
\end{figure*}
\addtocounter{figure}{-1}
\begin{figure*}[]
  \centering
  \includegraphics[width=19cm]{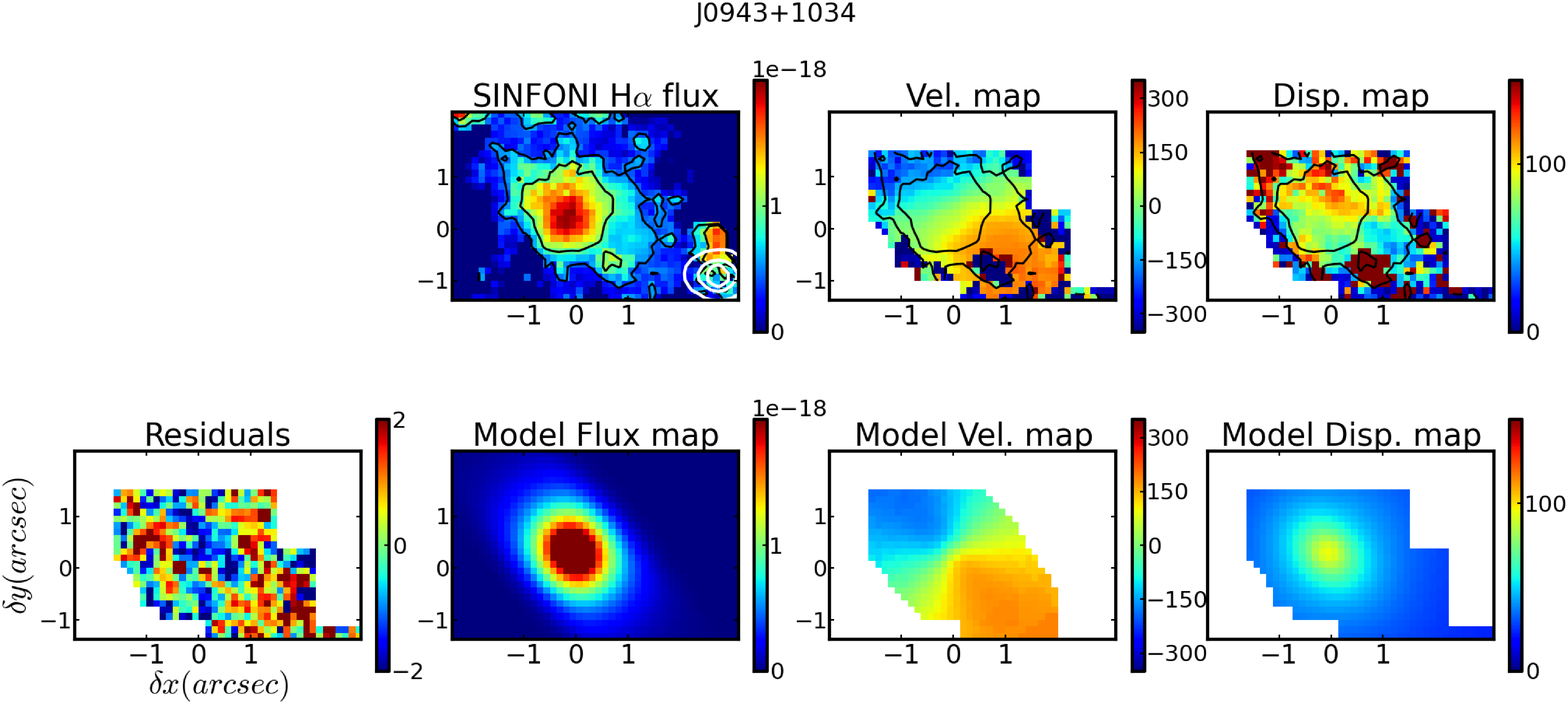}
  \includegraphics[width=19cm]{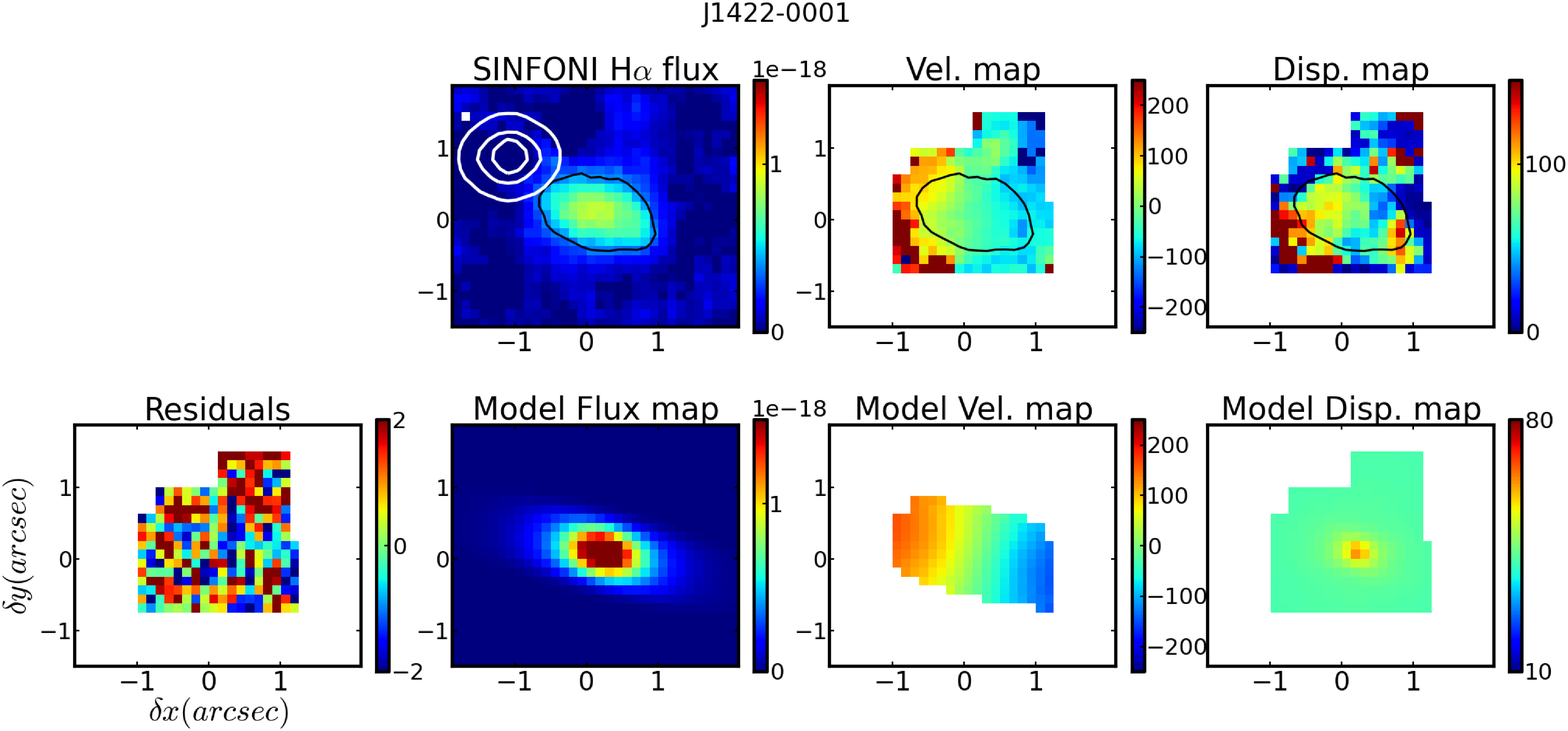}
  \includegraphics[width=19cm]{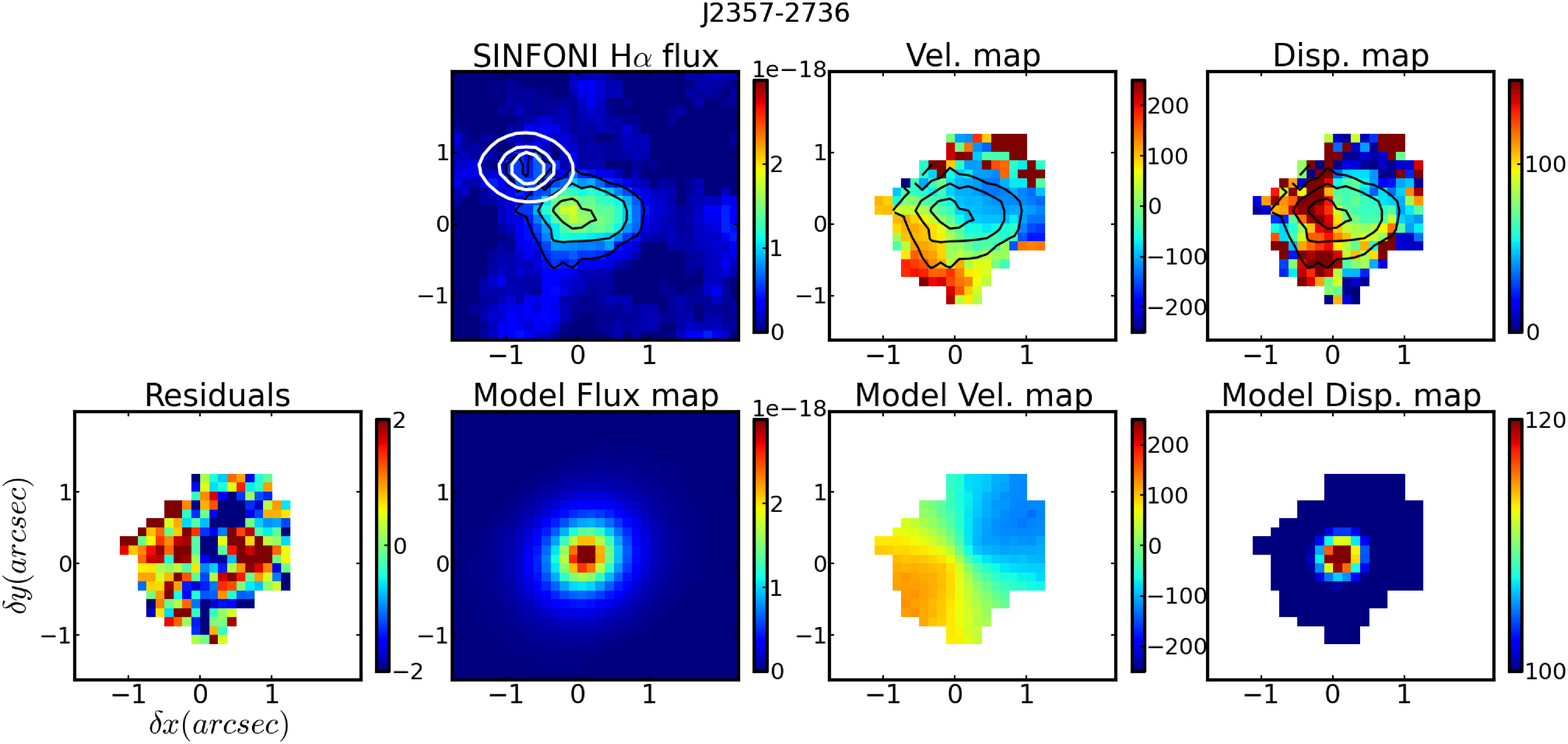}
  \caption{(continued)
  }
  \label{fig:flux_3}
\end{figure*}

\begin{figure*}[]
  \centering
  \includegraphics[width=19cm]{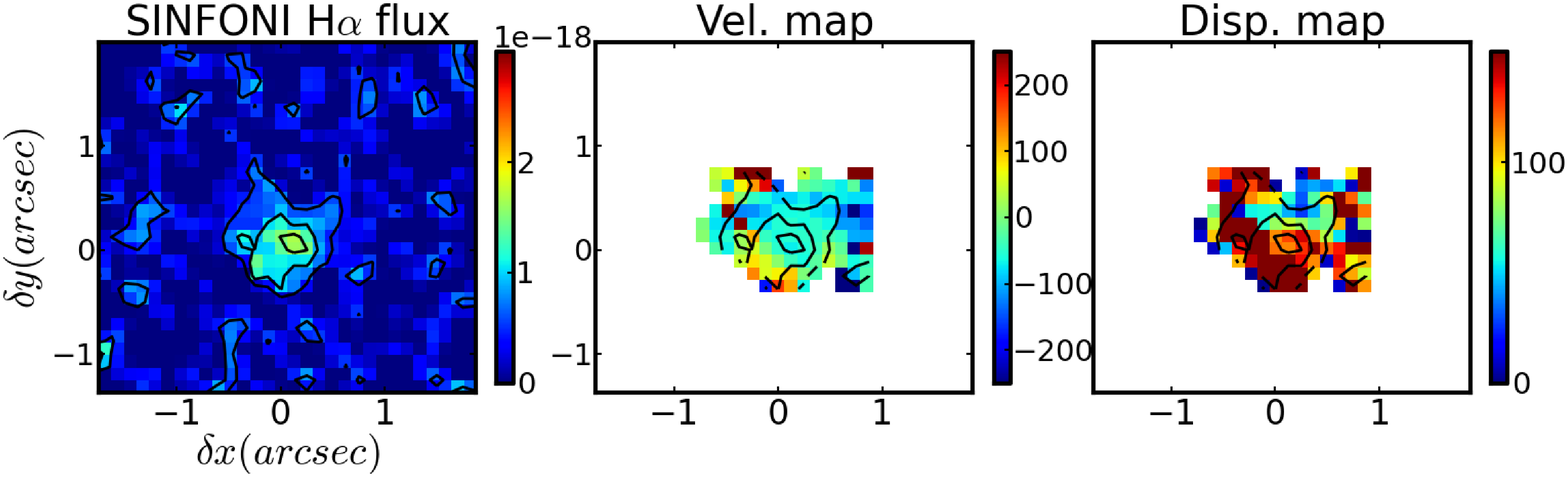}
  \caption{Same as Figure~\ref{fig:flux_1} but for the J1441+0443 galaxy.
  This galaxy has the lowest SNR, below the threshold where we can trust the  GalPaK$^{3D}$ results.
  Even if this galaxy is classified as \textbf{wind-pair }
  from its apparent PA, the low SNR
  does not allow us to build a wind model. 
  }
  \label{fig:flux_4}
\end{figure*}


\subsection{Redshifts}
\label{section:redshifts}

An accurate systemic redshift is crucial to characterize the outflow velocity and ultimately the outflow mass loading factor.
The GalPaK$^{3D}$ algorithm outputs the wavelength of the \Ha\ emission line from the axi-symmetric disk model. 
Since the galaxy distribution may be somewhat asymmetric, this sometimes lead to a redshift bias.

Therefore, we use two different methods. 
First, we determine the redshift from the mean of the wavelength of the reddest and the bluest (gaussian) 
\Ha\ emission lines along the kinematic major axis. 
\textbf{As a second check}, we create a pseudo-longslit along the kinematic major axis, and determine $z_{sys}$ from the sharp transition in the p-v diagram. 
We find that both methods yield consistent results.
These redshifts are listed in Table~\ref{table:galpak}.

\textbf{The resulting intrinsic galaxy} parameters will now allow us to build a cone model in order to reproduce the data for galactic outflows.


\subsection{Azimuthal dependence}
\label{section:azimuthal}

\begin{figure}[!h]
  \centering
  \includegraphics[width=8cm]{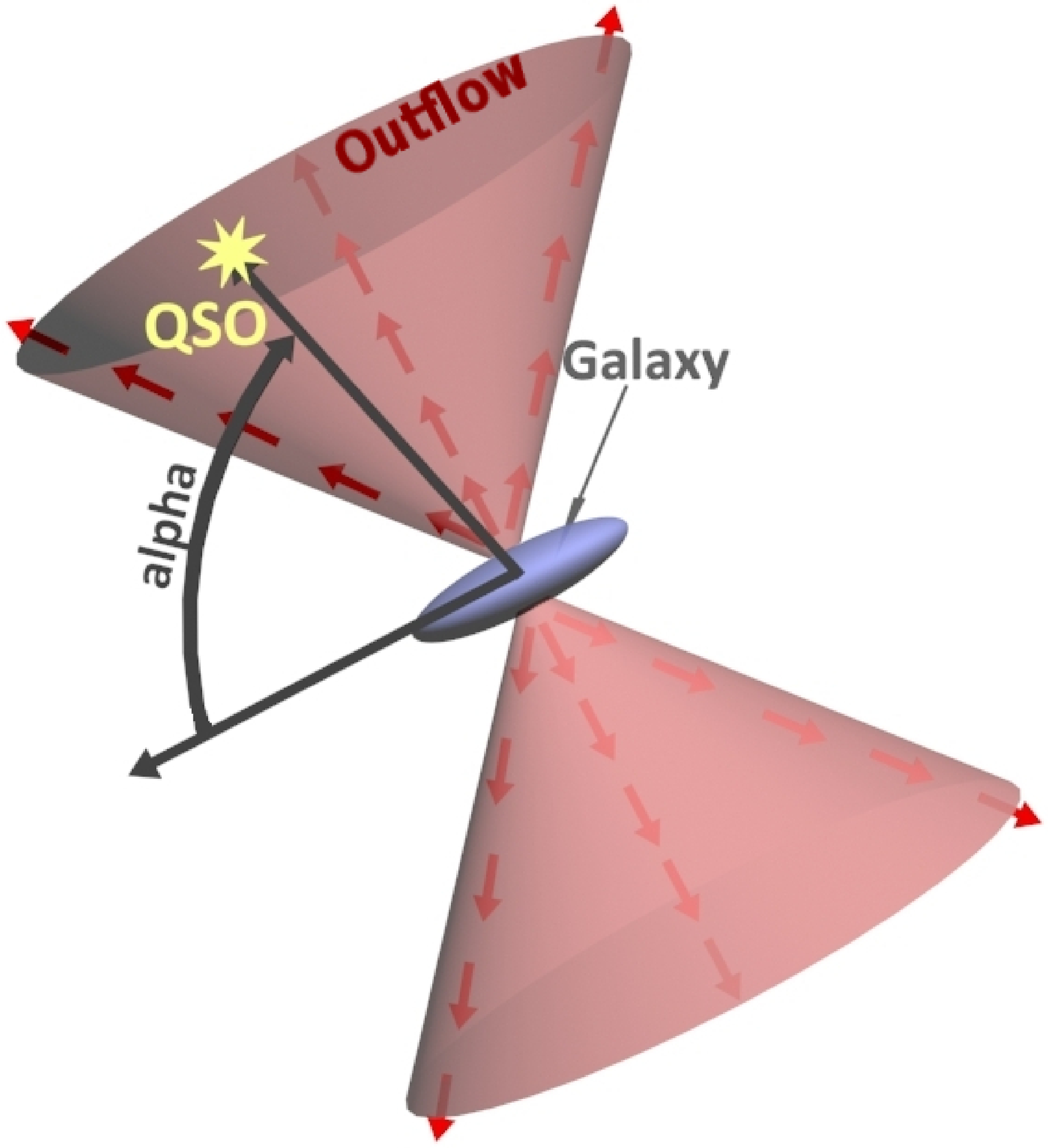}
  \caption{Scheme showing the alpha angle corresponding to the angle between the galaxy major axis and the quasar position.
  \label{fig:schema}
}
\end{figure}

In order to begin the wind modeling, we must first select galaxy-quasar pairs for which the quasar LOS  \textbf{intercepts }
the galactic winds.
This can be achieved using the quasar azimuthal angle $\alpha$   between the galaxy major axis and the quasar (Figure~\ref{fig:schema}),
because the presence of strong \MgII\ absorbers is a strong function of $\alpha$ as demonstrated by numerous recent studies \citep{bordoloi_11,bouche_12,bordoloi_14,lan_menard_14}
Hence, we use the quasar position relative to the associated galaxy major axis, using the inclination and major-axis determined from the SINFONI data (Figs~\ref{fig:flux_1}),
to classify the different galaxy-quasar pairs in two main categories (\textbf{pairs suitable for wind studies (wind-pair) }
for likely outflows and \textbf{pairs suitable for accretion studies (inflow-pair) }
for likely inflows).

\begin{figure*}[]
  \centering
  \includegraphics[width=15cm]{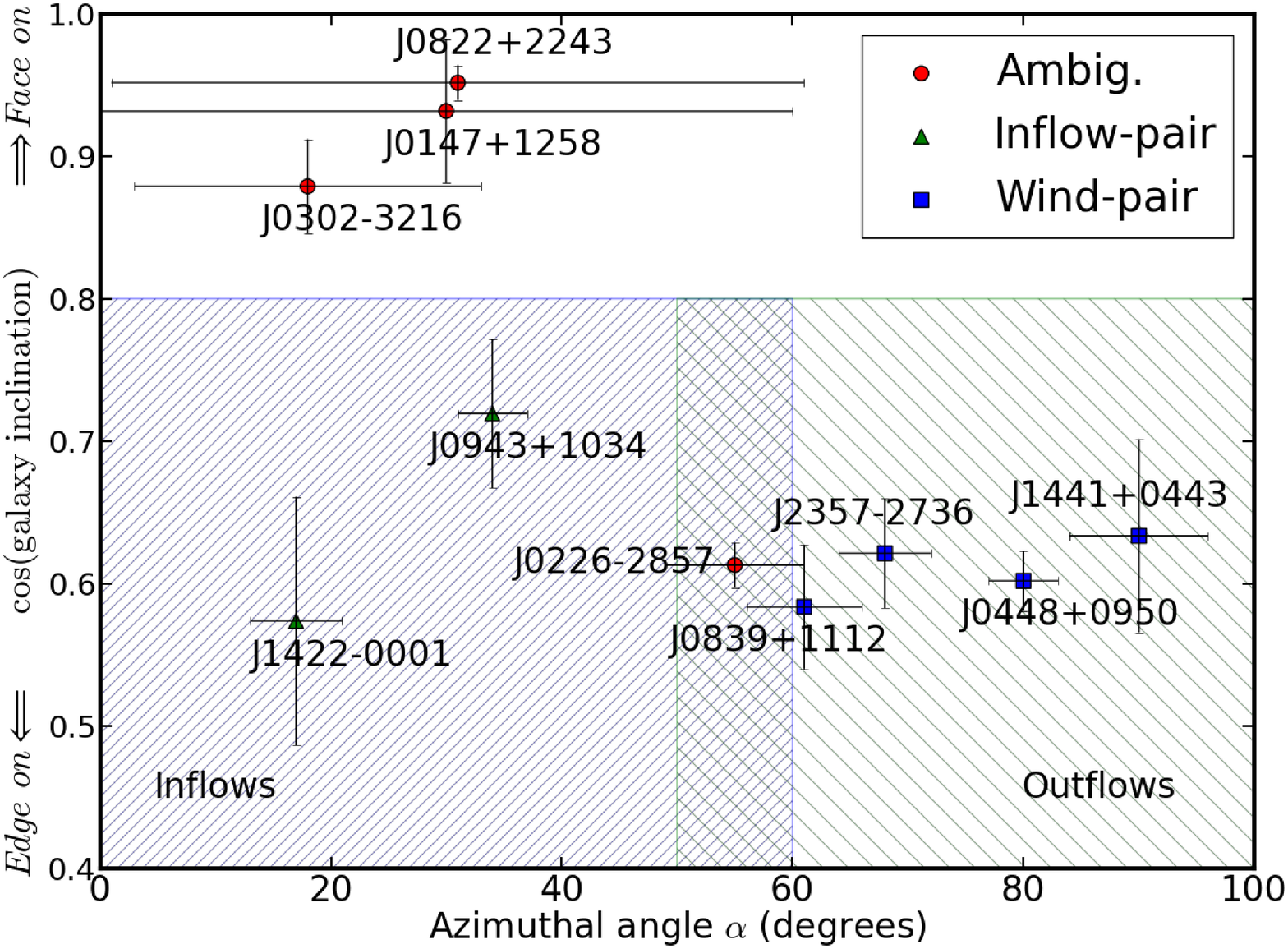}
  \caption{Galaxy inclinations for the SIMPLE sample as a function of the azimuthal angle $\alpha$.
  Note there are three types of galaxies in this sample: the \textbf{wind-pairs }
  which have an azimuthal angle larger than 60$\pm$10$^\circ$, the \textbf{inflow-pairs }
  with $\alpha$ lower than 60$\pm$10$^\circ$ 
  and pairs that are ambiguous due to uncertainty on $\alpha$. It is difficult to derive the azimuthal angle for a nearly face-on galaxy.
  The \textbf{wind-pair }
  and \textbf{inflow-pair }
  classes describe the fact of having the quasar absorptions tracing outflows and inflows, respectively. 
  \label{fig:alpha}
  }
\end{figure*}

Figure~\ref{fig:alpha} shows the azimuthal angles $\alpha$ versus the galaxy's inclinations for our SIMPLE sample of 10 galaxy-quasar pairs.
Pairs with $60^\circ \leq \alpha \leq 90^\circ$ are selected to be \textbf{wind-pairs}.
Pairs with $0^\circ \leq \alpha \leq 30^\circ$ correspond to the cases where the quasar LOS does not probe outflows but rather 
probes the extended parts of gaseous disks,  \textbf{where the gas can potentially (or is likely) to be inflowing (inflow-pair)} as in \citet{bouche_13}.

Naturally, the azimuthal angle for  galaxies with low inclinations, corresponding to face-on cases, is very difficult to constrain.
These are then indexed as 'face-on' cases. 
Pairs with $\alpha\sim 45^\circ$ correspond to ambiguous cases where it is  difficult to argue for outflows or inflows. For instance  J0226-2857 falls into that category
with the additional difficulty that this galaxy has a very low impact parameter ($b=0.3$ arcsec or $<$2.0 kpc), i.e. the line-of-sight is likely dominated by absorption from the galaxy interstellar medium.

Figure~\ref{fig:alpha} shows that four galaxies are favorable to study galactic winds properties: J0448+0950, J2357-2739, J0839+1112 and J1441+0443, and are 
classified as wind-$-$pairs in Table~\ref{table:galpak}.
J1441+0443 is excluded from subsequent analysis because our SINFONI data does not meet the requirement of SNR$\sim3$ imposed by our intensive tests of the GalPaK$^{3D}$ algorithm.
 
\section{Wind properties analysis}


\subsection{Wind sub-sample analysis}
\label{section:wind}

For each galaxy--quasar pair, we have the quasar spectrum for all the SIMPLE galaxies taken with the VLT/UVES instrument.
In these spectra, we identified 
\textbf{three} main absorption features: the \MgII\ (\ma\ \mb) doublet and the \MgI\ (\mc) absorption line.
Because of our selection in $W_r^{\lambda2796}$ of $2$ \AA\ (to ensure the host was in the SINFONI field-of-view), the \MgII\ doublet is saturated.
The \MgI\ absorption line is not saturated and in most cases shows an asymmetric profile.
This asymmetry can also be seen in the \MgII\ doublet, but less obviously.
We center the spectrum on every absorption line \textbf{using the derived redshifts. }
\textbf{For each absorption line, we transform wavelength to velocity}, taking the vacuum wavelengths.
\textbf{From the absorption system kinematics and geometrical properties of the galaxy, }
we can now build the winds model for the three galaxies.


\subsubsection{Cone wind-modeling}
\label{section:cone}

We follow \citet{bouche_12} and \citet{kacprzak_14} in modeling the wind as a bi-conical outflow using the geometric parameters (inclination, $\alpha$) set by the SINFONI data.
The principle is to create a cone perpendicular to the galactic plane, fill it with uniformly distributed particles and assume that the mass flux is conserved.
The  particles represent cold gas clouds entrained in the wind, since
the equivalent width of \textbf{the} absorption lines is the sum/combination of several saturated lines \citep{menard_09}, each of which corresponds to a `cold' 
gas cloud ($10^4K$) entrained by supernovae \textbf{heated} hot winds ($T>10^6K$).
\textbf{Since the galaxy inclination, }PA and azimuthal angle are previously determined from observations using GalPaK$^{3D}$ and other methods, 
the only free parameters are $V_{\rm out}$ and the cone opening angle $\theta_{\rm max}$ (see the effect of both in the Appendix).
For simplicity, we \textbf{assign the clouds } 
a constant radial velocity $V_{\rm out}$, i.e. we assume that the \textbf{LOS }
intercepts the clouds far from the acceleration region.

The cone is built along the x, y, z axes: \textbf{x and y represent the sky plane and z corresponds to the cone height.}
For a galaxy with 0$^\circ$ inclination, the cone direction will be along the line of sight.
We then rotate the cone along the y-axis to match the galaxy's inclination derived from our SINFONI data and create a simulated absorption profile from
the distribution of cloud velocities projected along the quasar LOS (z axis).

We generate $\sim$10$^6$ particles in a cone\textbf{, which are grouped by }
bins of projected velocities.
The quasar LOS is set by the impact parameter ($b$) and $\alpha$, both of which are 
derived from the SINFONI data cubes. 
Due to the Monte Carlo generation of particles, \textbf{stochastic effects create noise in the simulated profiles. }
This noise does not impact the resulting equivalent widths and thus the derived outflow velocities.

We then convolve the \textbf{particle } velocity distribution with the UVES instrument resolution.
\textbf{Additionally, in order to simulate the instrument noise, we add a random Poisson noise to the simulated profile.
This random Poisson noise has the same signal to noise ratio as the data and provides for a more meaningful comparison. }
In order to give an intuitive feel for this geometric model, we show in the Appendix examples of simulated profiles using different galaxy inclinations, outflow velocities and opening angles.


\subsubsection{Galaxy contribution model}

Since our sample \textbf{consists} of pairs with small impact parameters ($b < $ 20~kpc) and with inclined galaxies (from $\sim 18^{\circ}$ to $\sim 55^{\circ}$), 
 we improve our model by adding the galaxy contribution for the quasar-galaxy pairs with 
the lowest impact parameters ($\rm b \leq 10$ kpc) such as J2357-2736 \textbf{(Section~\ref{section:accelerated})}. 
The procedure is nearly the same as the cone model: we generate particles in a disk with an exponential distribution from the center to the edge.
We take the galaxy half light radius derived with GalPaK$^{3D}$ \textbf{to estimate a} realistic contribution from the disk.
The thickness of the disk is set to be 0.15 times this radius.
\textbf{We assign the particles} a constant circular velocity corresponding to the maximum velocity of the galaxy. 
The velocity distribution of the disk is naturally strongly dependent on the azimuthal angle with a maximum offset at $\alpha=0^{\circ}$  and a distribution centered
around 0 \kms at $\alpha=90^{\circ}$.

\subsection{Comments on individual wind-pairs}

\subsubsection{J0448+0950}

The galaxy near the quasar J0448+0950 has an impact parameter $b=$ 13.7 kpc and an \Ha\ flux of $5.03\times 10^{-16}$ \flux.
Its azimuthal angle $\alpha$ of $\sim$79$^\circ$ and inclination $i$ of $\sim$52$^\circ$ make it a \textbf{wind-pair }
(Figure~\ref{fig:alpha}).
This galaxy has a SFR of $\sim$13 \mpy\ (see section 4) and a redshift of 0.8390.

In addition to our SINFONI data, we retrieved ancillary data from HST/WFPC2 (F555W filter). 
These HST data allow us to compare the morphology of the galaxy (see Figure~\ref{fig:flux_2}) with the SINFONI one. 
In both data sets, one sees that the galaxy has an asymmetric flux distribution (Figure~\ref{fig:flux_2})
with a brighter area  somewhat offset with respect to the kinematic center.
Comparing \textbf{the} HST image and SINFONI flux map (the quasar was subtracted in SINFONI \Ha\ flux), the PA and inclination of the galaxy are in  good agreement.

\begin{figure*}[]
  \centering
  \includegraphics[width=15cm]{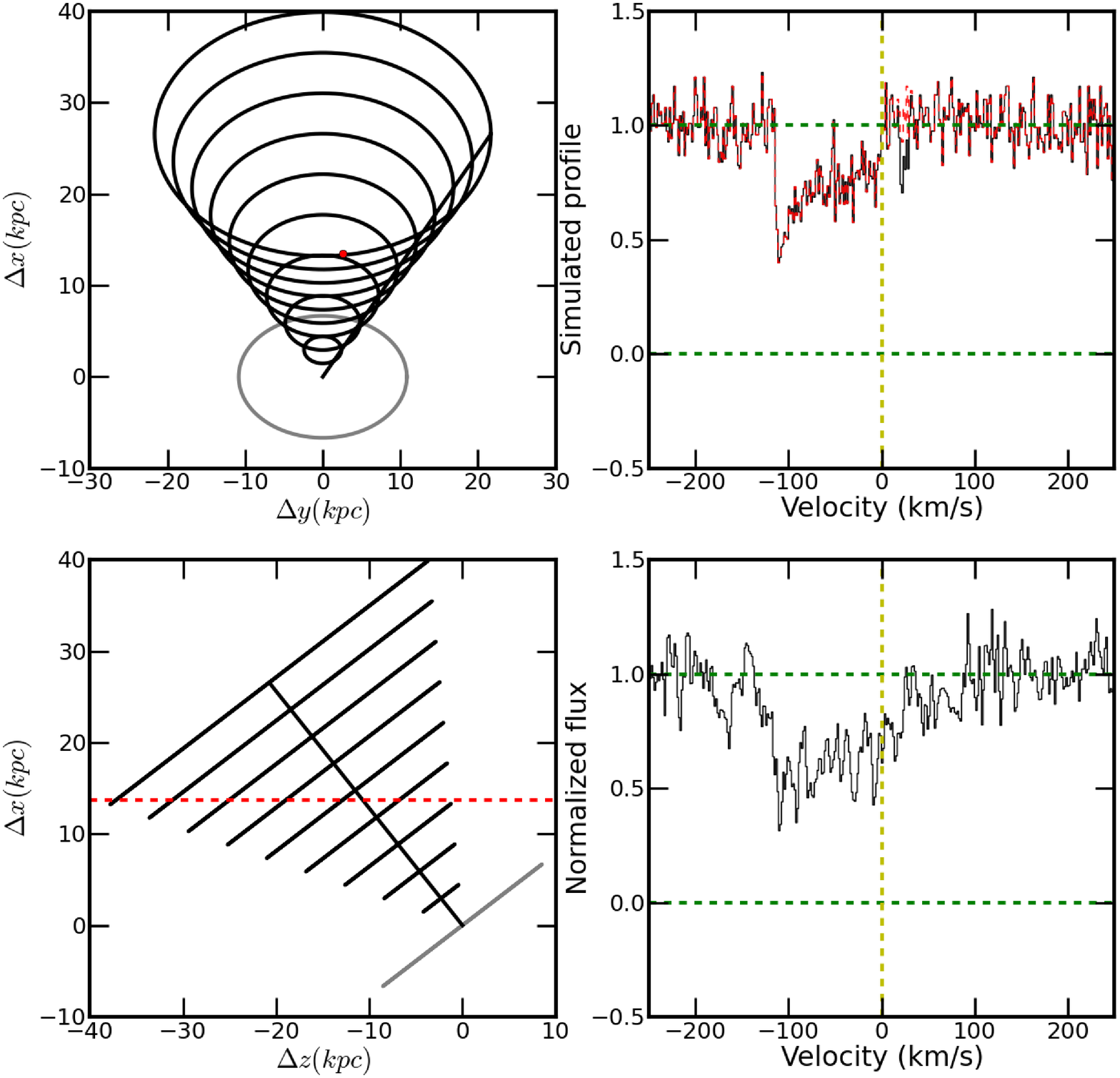}
  \caption{
  Representation of the cone model and quasar spectrum associated with the J0448+0950 galaxy. 
  \textit{Left: }the cone model seen in the sky plane where the y axis corresponds to the galaxy major axis and x to its minor one.
  The gray circle represents the inclined galaxy disk and the black circles \textbf{illustrate} the gas outflow cone.
  \textit{Bottom left: }a side view of the cone where the z-axis corresponds to the quasar LOS direction with the observer to the left.
  \textit{Right: }Normalized flux for the \MgI\ \mc\ absorption line \textbf{observed with UVES }(bottom) \textbf{where we can see an outward asymmetry}, and the 
  \textbf{reconstructed} profile (top).
  The red dashed \textbf{line gives} the simulated profile without taking into account the galaxy contribution.
  The black line does take into account this contribution.
  Note that this model does not reproduce the depth of the absorption line.}
  \label{fig:J0448}
\end{figure*}

\textbf{After determining} the geometrical parameters for this galaxy, we can build a cone model as described in section~\ref{section:cone}.
In Figure~\ref{fig:J0448}, we compare the simulated profile for  \MgI\ \mc~\footnote{We use \MgI\ \mc\ since \MgII\ \ma\ is saturated.
.} \textbf{to the observed absorption in the UVES data }
(right column of the figure).
To generate this simulated profile~\footnote{The galaxy contribution is also considered in the simulated profile (the redshifted contribution in upper right of Figure~\ref{fig:J0448}).}, we adjusted the outflow speed $V_{\rm out}$ and the cone opening angle $\theta$, while keeping the geometrical parameters fixed.
The best values are an outflow speed $V_{\rm out}$ of 115$\pm$10 \kms\ and a cone opening angle $\theta_{\rm max}$ of 40$\pm$5$^\circ$.
The errors represent the maximum allowed range values for $V_{\rm out}$ and $\theta_{\rm max}$.

We note that our simulated profile \textbf{reproduces the asymmetry and equivalent width of the observed profile.}
Note, our model does not attempt to reproduce the depth of the profile since it is arbitrarily normalized.
The apparent noise of the simulated profile is due to stochastic effects from \textbf{the }Monte Carlo particle distribution.

Outflow rates and mass loading factors for each galaxy identified as \textbf{wind-pairs} are detailed in section~\ref{section:loading_factor}.

\subsubsection{J0839+1112}

In our sample, the \textbf{galaxy towards} J0839+1112 has the largest impact parameter of $b$ of 26.8 kpc.
With an \Ha\ flux of $1.53\times10^{-16}$ \flux, an inclination $i$ of $\sim$72$^\circ$ and an azimuthal angle $\alpha$ of $\sim$59$^\circ$, this galaxy also belongs to the \textbf{wind-pair} 
subsample defined in section~\ref{section:azimuthal}.
Its SFR is $\sim$3.4 \mpy\ , and \textbf{it} has a redshift of $z=$0.7866.

In Figure~\ref{fig:flux_2} we compare archival HST/WFPC2 (F702W filter) 
images to our SINFONI \Ha\ data.
Both data sets show a slight asymmetry in the galaxy flux distribution and a similar PA.
For the galaxy inclination,  we used galfit2D on the HST image to cross-check the results from the 3D fitting
and found good agreement between them \textbf{and found that the results were within} $\pm 15^\circ$.

\begin{figure*}[]
  \centering
  \includegraphics[width=15cm]{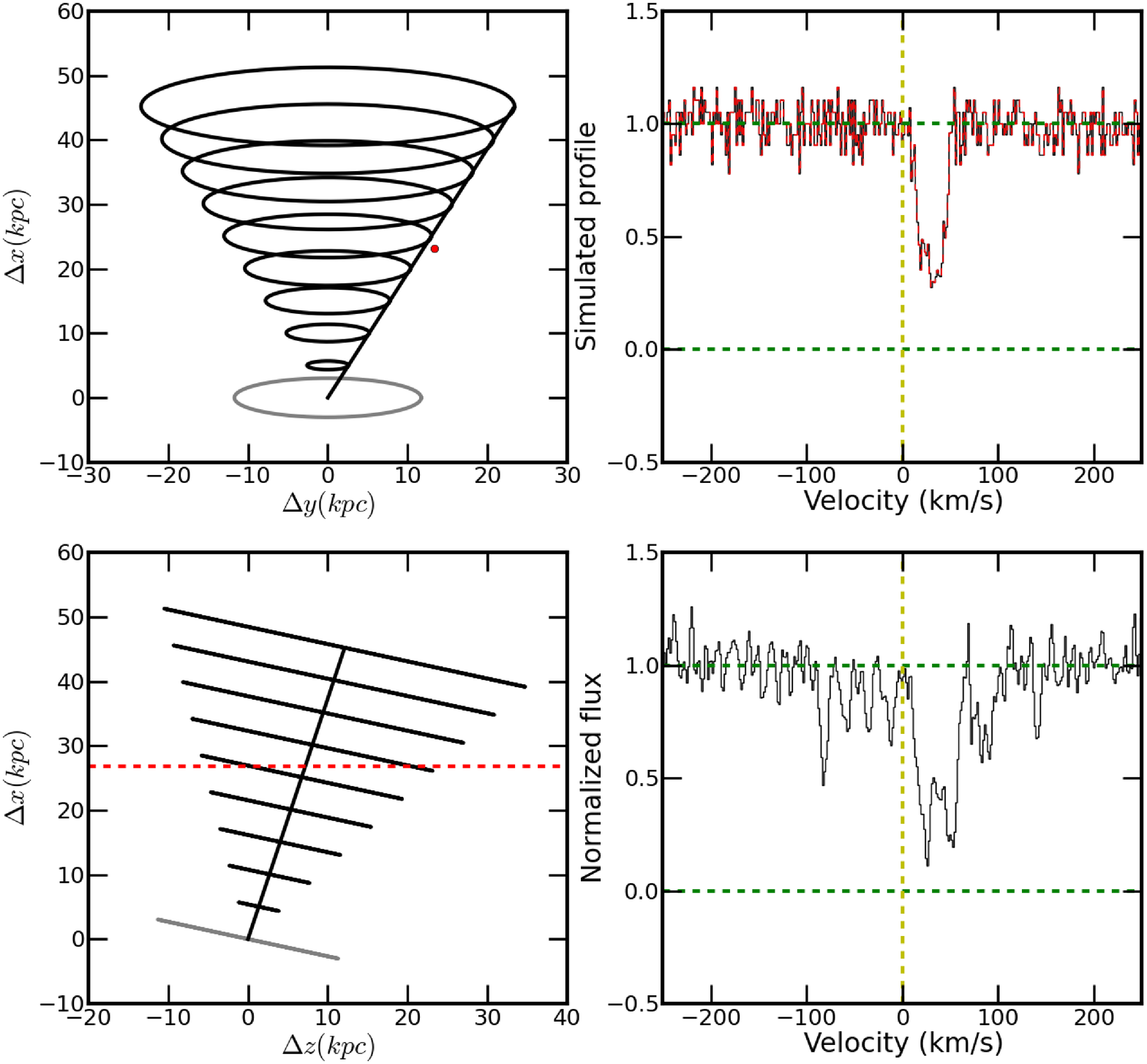}
  \caption{Same as Figure~\ref{fig:J0448} for the J0839+1112 galaxy \textbf{where no galaxy contribution can be seen}.}
  \label{fig:J0839}
\end{figure*}

As for J0448$+$1112,  we generated a simulated profile from the wind cone model using the geometrical parameters from the SINFONI$+$HST data and  
adjusted the outflow speed $V_{\rm out}$ and the cone opening angle $\theta$, while keeping the geometrical parameters fixed. Figure~\ref{fig:J0839} shows 
the simulated profile and the \MgI\ \textbf{absorption} from the UVES data (right column of the figure).
We \textbf{constrained} an outflow speed $V_{\rm out}$ of 105$\pm$10 \kms and a cone opening angle of $\theta_{\rm max}$ of 30$\pm$5$^\circ$. 
The impact parameter $b$ is too high to consider any contribution from the galaxy.

\subsubsection{J2357$-$2736}
\label{section:accelerated}

The last individual case from our wind subsample is the galaxy along the J2357$-$2736 LOS.
The host galaxy has the smallest impact parameter $b$ to the quasar LOS with $b$ of 6.7 kpc.
This galaxy has an \Ha\ flux of $1.29\cdot10^{-16}$ \flux\ and a SFR of $\sim$3.3 \mpy\ .
Its inclination $i$ is $\sim$51.6$^\circ$ and it is classified as \textbf{wind-pair} 
because of its azimuthal angle $\alpha$ of $\sim$68$^\circ$.

\begin{figure*}[]
  \centering
  \includegraphics[width=14cm]{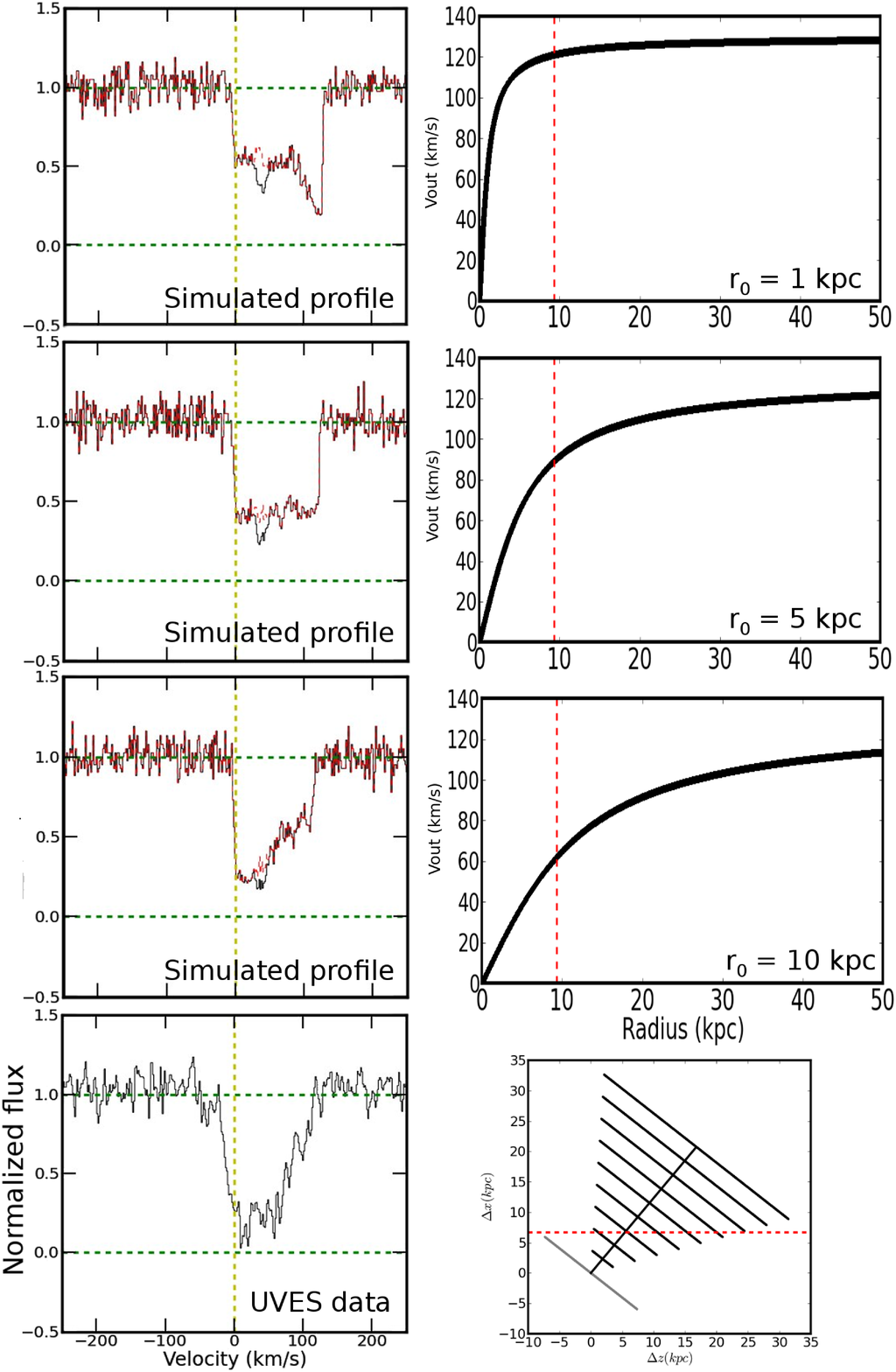}
  \caption{
  \textit{Left column: }\textbf{\MgI\ absorption from the J2357$-$2736 galaxy indicate that the model must include accelerating winds.}, 
  from top to bottom: simulated absorption profiles with $r_0$=1,5,10 kpc, and the UVES data centered on \MgI\ \mc\ .
  \textbf{Notice that the asymmetry changes as $r_0$ increases, it goes from outward to inward asymmetry.} 
  \textit{Right column:} The velocity profile corresponding to the associated simulated profile to the left \textbf{where the turn over radius of the velocity profile ($r_0$) varies}, 
  from top to bottom: $r_0$=1,5,10 kpc. 
  The red dashed line represents the distance between the galaxy 
  and the quasar LOS ($b$/sin($\alpha$)/sin($i$)), corrected for the inclination $i$.
  The final simulated profile is the one \textbf{directly} above the data, with $r_0$=10 kpc.}
  \label{fig:J2357}
\end{figure*}

\begin{figure}[]
  \centering
  \includegraphics[width=8cm]{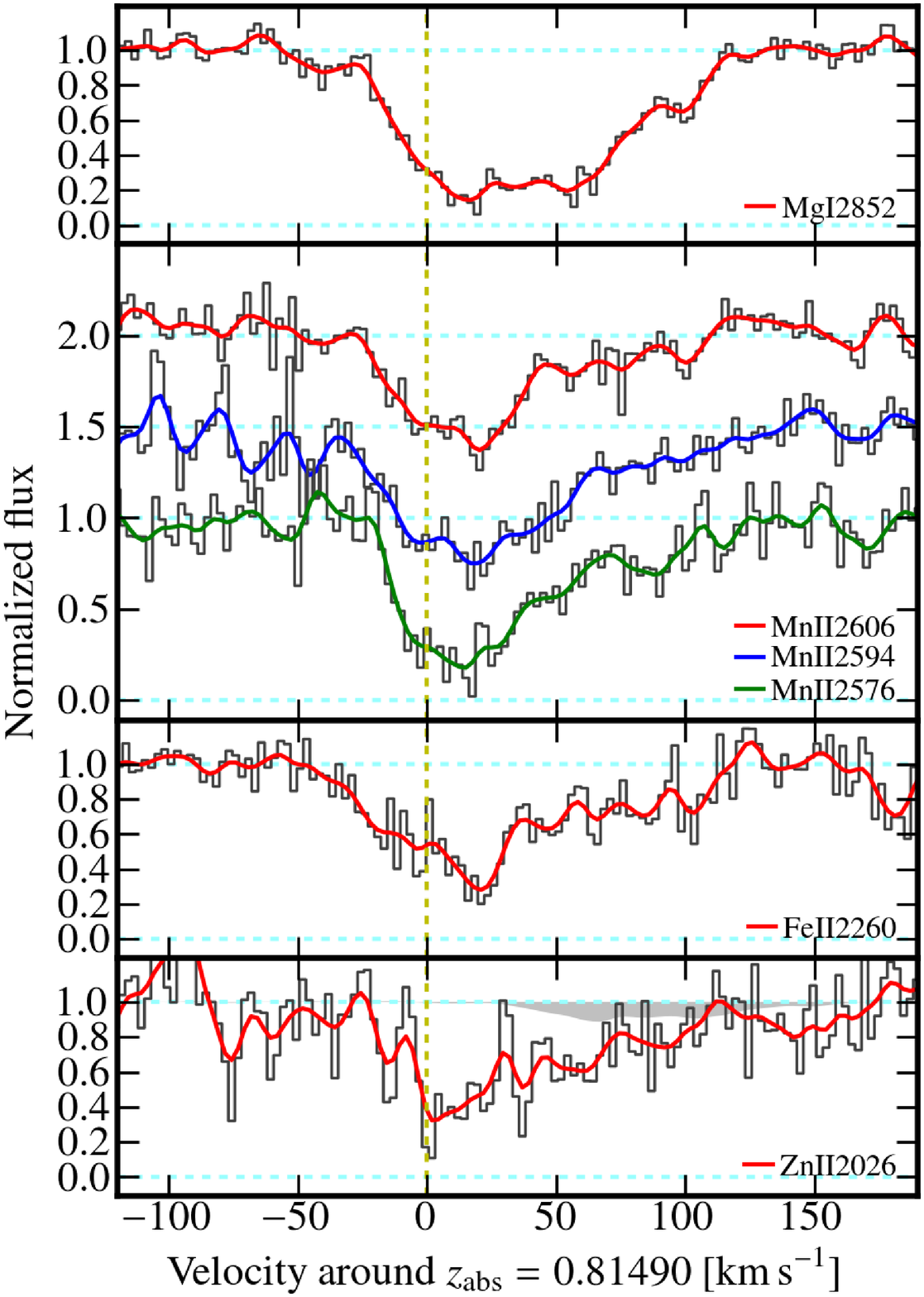}
  \caption{Absorption lines \textbf{observed with UVES} for J2357-2736. The solid lines are a Gaussian smoothing of the data (with
a sigma of 1 pixel) to aid the eye. The grey shaded area in the \ZnII (\zna )
panel is the absorption expected from the \MgI (\zna ) line (derived by scaling
the observed optical depth of the \MgI (\mc ) line, ignoring potential
saturation effects).
  }
  \label{fig:J2357_lines}
\end{figure}

As \textbf{in} the previous two cases, we generated a simulated UVES profile using the wind model described in section~\ref{section:wind}.
Figure~\ref{fig:J2357} (bottom) shows the  UVES \MgI\ \mc\ absorption profile, whose asymmetry is reversed compared to the two other cases with a maximum optical depth at $V\sim0$ \kms.
However, any constant wind speed model
will have an outward asymmetry (Figure~\ref{fig:models}) and the data clearly shows the opposite, an inward asymmetry.
\textbf{For a profile with inward asymmetry, the strongest part of the absorption profile is located closer to the systemic velocity (e.g. bottom left of Figure~\ref{fig:J2357}).
An outward asymmetry profile has the opposite behavior (e.g. Figure~\ref{fig:J0448}).}
This inward asymmetry is seen in the other non-saturated transitions (\ZnII (\zna ), \MgI (\zna ), \MnII (\mniic, \mniib, \mniia))
present in the UVES data (Fig.~\ref{fig:J2357_lines}).

Contrary to the other two cases, this galaxy has a very low impact parameter ($b \sim 6.7 \rm kpc$),
where the assumption of constant wind speed might break down.
Indeed, the low-ionization material in momentum-driven winds and energy-driven winds is thought to be accelerated \citep[e.g.][]{murray_05, steidel_10} 
by the hot gas, by the radiation pressure, or both. 

Therefore, instead of using a constant wind speed, we added a generic velocity profile  such as $V(r)=V_{\rm out} \; 2 / \pi \; \arctan(r / r_0)$  
\textbf{where} $r$ is the distance from the galaxy and $r_0$ is the characteristic turn-over radius.
Figure~\ref{fig:J2357} shows the behavior of this model (dashed line) on the profile asymmetry for different values of $r_0$, illustrating that the asymmetry reverses as $r_0$ increases.
The accelerated wind model that best describes the data is the one with $r_0=10$~kpc shown above the UVES spectrum in Figure~\ref{fig:J2357}. 
\textbf{Similar to} J0448+0950, we also included a contribution from the galaxy which appears to account for the bluest components.

For this case, we found an outflow speed $V_{\rm out}$ of 130$\pm$10\kms using a cone opening angle $\theta_{\rm max}$ of 45$\pm$5$^\circ$.
Note that there is no degeneracy between $V_{\rm out}$ and $r_0$ as the various simulated profiles shown in Figure~\ref{fig:J2357} 
are for the same outflow velocity. In other words, $V_{\rm out}$ is set by the reddest part of the profile, \textbf{whereas} $r_0$ is constrained by the profile shape.

\textbf{Having determined} outflow velocities for the three \textbf{wind-pairs}
, we now focus on deriving the ejected outflow rates $\dot M_{\rm out}$ together with the mass mass loading factors $\eta$.

\subsection{Outflow Rate}
\label{section:loading_factor}
For each galaxy, the equivalent width of the \MgI\ \mc\ absorption lines only depends on $\theta_{\rm max}$ and $V_{\rm out}$: 
since every particle has the same velocity \textbf{or accelerated velocities up to $V_{\rm out}$ 
(see Section~\ref{section:accelerated})}, the projected velocity depends on the particle position in the LOS.
The equivalent width is due to the accumulation of the particles projected velocities.
We tested several velocities and opening angles in order to fit this width.
The profile asymmetry depends of the system geometry, the particle density and the outflow velocity ($V_{\rm out}$).
Quasar LOS that cross the wind cone \textbf{can constrain} the ejected mass rate \textbf{according to} relation (1) from \citet{bouche_12}:\\

\begin{eqnarray}
 \dot M_{\rm out} &\approx& \mu \cdot N_{\rm H}(b) \cdot b \cdot V_{\rm out} \cdot \frac{\pi}{2} \cdot \theta_{\rm max}\label{eq:Mout}\\
{\dot M_{\rm out} \over 0.5\/ \mpy\ }& \approx &{\mu \over 1.5} \cdot {N_{\rm H}(b)\over 10^{19} \rm cm^{-2}} \cdot {b\over 25 \rm kpc} \cdot {V_{\rm out} \over 200 \kms\ } \cdot {\theta_{\rm max} \over 30^{\circ}}\nn
\end{eqnarray}

$\mu$ being the mean atomic weight, $b$ the impact parameter, $\theta_{\rm max}$ the cone opening angle\footnote{\textbf{$\theta_{\rm max}$ is defined from the central axis, 
and the cone subtends an area $\Sigma$ of $\pi \cdot \theta_{\rm max}^2$.}} 
and $N_{\rm H}(b)$ 
is the gas column density of hydrogen at the $b$ distance.

The only parameter that remains to be constrained is the column density $N_{\rm H}(b)$.
In order to determine the gas column density $N_{\rm H}$, we use the empirical relation of \citet{menard_09}
between neutral gas column \textbf{density} and \MgII\ equivalent width $W_r^{\lambda 2796}$:
\begin{equation}
 N_{\HI} = \log [(3.06 \pm 0.55) \times 10^{19} \times (W_r^{\lambda 2796} )^{1.7 \pm 0.26}]\label{eq:menard}.
\end{equation}
This relation together with the tight correlation between \MgII\ equivalent width and
dust content (as determined statistically from quasar extinction) \textbf{from} \citet{menard_09}
 leads to a gas-to-dust ratio slightly smaller than that of the Milky Way \HI\ \textbf{column densities} of $\log (N_{\HI})=$19.5 and above.  
 \textbf{Furthermore, the redshift evolution of the dust content of \MgII\ absorbers extrapolated to $z=0$ shows that \MgII--selected aborbers extend the local  relation between
visual extinction $A_V$ and the total hydrogen column $N_{\rm H}$ of \citet{Bohlin_78}.}
This in turn \textbf{indicates} that  the ionized gas contribution is negligible in regions with \HI\ columns above $\log (N_{\HI})=$19.5
, as also argued by \citet{jenkins_09}, and that one can use the correlation between \MgII\ equivalent width and $N_{\HI}$ as a proxy for the $N_{\rm H}$ gas column density.

Given our selection criteria of $W_r^{\lambda 2796}>2$~\AA, we are very likely in a regime where the gas is mostly neutral.  
\textbf{For our three wind-pairs sight lines, the \HI\ column densities are: $\log (N_{\HI})\approx20.3$ for J0448+0950, $\log (N_{\HI})\approx20.1$ for J0839+1112 and $\log (N_{\HI})\approx19.9$ for J2357$-$2736.
The rest equivalent widths $W_r^{\lambda2796}$ determined from the UVES data and the corresponding $N_{\rm H}$ column densities for the wind-pair galaxies are listed in Table~\ref{table:windresults}.
In future work, we will be able to measure $N_{\HI}$ directly
from UV spectroscopy with {\it HST}/COS.}

\begin{figure}[]
  \centering
  \includegraphics[width=9.0cm]{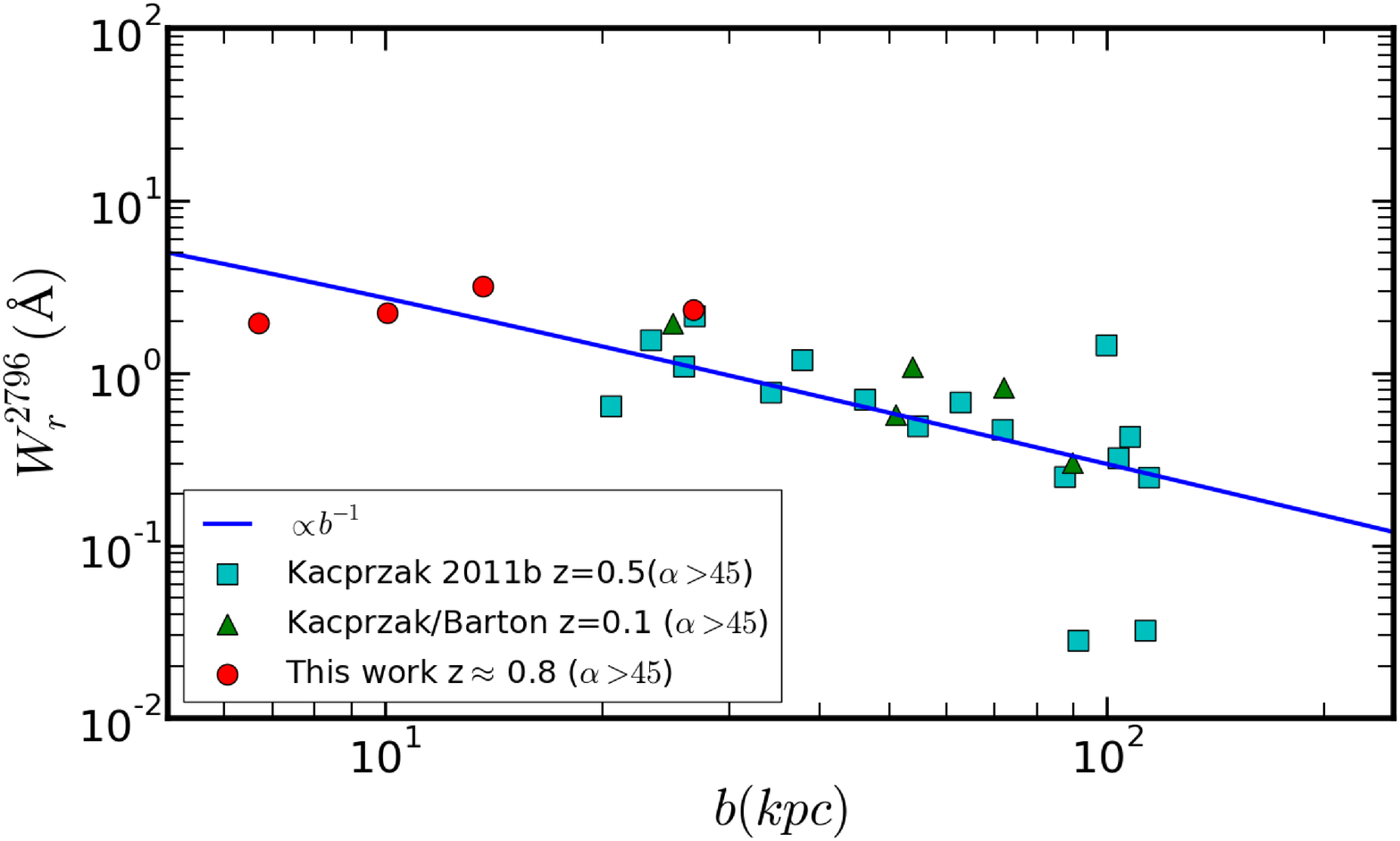}
  \caption{$W_r^{\lambda 2796}$ as a function of impact parameter $b$ for galaxy-quasar pairs classified as \textbf{wind-pairs}.
  }
  \label{fig:REW_vs_b}
\end{figure}

Figure~\ref{fig:REW_vs_b} shows the \MgII\ rest equivalent width $W_r^{\lambda2796}$ as a function of impact parameter $b$
 for quasar-galaxy pairs where the quasar is aligned with the minor axis (\textbf{wind-pairs})
 from various literature samples \citep[][]{kacprzak_11, kacprzak_11b} and this paper.
This figure shows that the tight anti-correlation between impact parameter $b$ and $W_r$ (\citet{bouche_12}) is confirmed at $b<30$ kpc.
The solid line traces the fiducial $1/b$ relation for mass-conserved bi-conical outflows \citep[see][]{bouche_12}.

\textbf{Table ~\ref{table:windresults} lists our estimated outflow rates determined using Eq.~\ref{eq:Mout}. 
In order to determine the error bars for $\dot M_{\rm out}$, we take the maximum error of every parameter used to derive it. 
We thereby objectively determine the maximum uncertainty on the ejected mass rate.}
\textbf{We also note that the errors on $\dot M_{\rm out}$ are dominated by the errors on $N_{\HI}$.}

\textbf{From the} outflow rates, we compute the mass mass loading factor $\eta$ by comparing it to the SFR.
\textbf{We derived the SFR from \Ha\ }
using the \citet{kennicutt_98} calibration, which assumes a \citet{salpeter_55} Initial Mass Function (IMF):

\begin{equation}
 SFR (\rm{M_{\odot}~yr^{-1}})=7.9\times 10^{-42}~ L_{H\alpha} \label{eq:sfr}
\end{equation}
where $L_{H\alpha}$ is the \Ha\ luminosity in \flux.
We note \textbf{that} the SFRs for the Salpeter IMF with no extinction correction are identical  to  using a dust correction of 1 mag \citep{zahid_13} with the \citet{chabrier_03} IMF,
as the two IMFs are offset by -0.25 dex (see Table 2 in \citet{bernardi_10}).

The results for the three galaxies are shown in Table~\ref{table:windresults}.
If we assume that galactic winds are symmetric with respect to the galactic plane (Figure~\ref{fig:schema}), 
the total ejected mass rate for a galaxy must be increased by a factor of 2, which gives $\dot M_{\rm out} \approx$ 9 \mpy\ for J0448+0950, $\dot M_{\rm out} \approx$  2 \mpy\ 
for J2357-2736 and $\dot M_{\rm out} \approx$ 7 \mpy\ for J0839+1112.\\

\textbf{Considering} the ejection velocity of the winds \textbf{(115, 105 and 130 \kms\ for J0448+0950, J0839+1112 and J2357$-$2736 respectively)}, it is interesting to test whether this 
velocity is large enough for the gas to leave the galaxy halo or if it will end up falling back onto the galaxy.
The escape velocity $V_{\rm esc}$ for an isothermal sphere is given by the following Eq.~\ref{eq:Vesc} \citep{veilleux_05}:
\begin{equation}
 V_{\rm esc}=V_{\rm max}\cdot\sqrt{2\left[1+\ln \left(\frac{R_{\rm vir}}{r}\right)\right]}\label{eq:Vesc}
\end{equation}
where $V_{\rm max}$ is the maximum rotation velocity of the galaxy and $R_{\rm vir}$ is the virial radius.
Given that our galaxies halos have a mass close to 10$^{12}$ \msun\ , their virial radius is approximately $R_{\rm vir} \approx V_{\rm max}/10H(z)$ where $H(z)$ is the Hubble constant at redshift $z$.
\textbf{For the wind-pairs, their virial radii are 225 kpc for J0448+0950, 103 kpc for J0839+1112 and 168 kpc for J2357$-$2736.}
\textbf{We give these results in Table~\ref{table:windresults}, along with }
the results on mass loading factors $\eta = \dot M_{\rm out}$/SFR.

\begin{table*}[h]
\centering
\caption{Results for galaxies J0448+0950, J2357-2736, J0839+1112 together with literature results.}
\label{table:windresults}
\begin{tabular}{lcccccccccc}
\hline
Galaxy         & $b$ (kpc) & log($N_{\rm H}(b)$) & $V_{\rm max}$        & $V_{\rm out}$ (\kms )   &$\theta_{\rm max}$    &   SFR                 & $\dot M_{\rm out}$        & $\frac{V_{\rm out}}{V_{\rm esc}}$ & $\eta$ & Reference\\
(1)            & (2)     & (3)                 & (4)                  & (5)                     & (6)                  & (7)         & (8)       & (9)              & (10)       & (11)          \\
\hline
J0448+0950     & 13.7    & 20.30$\pm$0.3         & $253 \pm 10$         & $115 \pm 10$            &  40 $\pm$ 5.0        &   13.6$\pm$0.3        &  4.6 $^{+4.9}_{-3.2}$          &   0.16     &   0.70    & This work     \\
J0839+1112     & 26.8    & 20.10$\pm$0.3         & $115\pm8$            & $105  \pm 10$           &  30 $\pm$ 5.0        &   3.4$\pm$0.2         &  3.6 $^{+3.4}_{-2.2}$          &   0.43     &   2.11    & This work     \\
J2357$-$2736   & 6.7     & 19.92$\pm$0.2         & $186 \pm 15$         & $130 \pm 10$            &  45 $\pm$ 5.0        &   3.3$\pm$0.2         &  1.2 $^{+1.1}_{-0.7}$          &   0.24     &   0.75    & This work     \\
J081420G1      & 51.1    & 19.07$\pm$0.2         & $131 \pm 10$         & $175 \pm 25$            &  30 $\pm$ 5.0        &   5.0$^\dagger$       &  1.0 $^{+1.4}_{-0.7}$          &   0.63     &   0.42    & B2012        \\
J091119G1      & 71.2    & 19.34$\pm$0.2         & $231 \pm 10$         & $500 \pm 100$           &  30 $\pm$ 5.0        &   1.2$^\dagger$       &  7.8 $^{+12.2}_{-4.5}$         &   0.97     &   12.9    & B2012        \\
J102847G1      & 89.8    & 18.60$\pm$0.2         & $162 \pm 10$         & $300 \pm 25$            &  30 $\pm$ 5.0        &   9.0$^\dagger$       &  1.1 $^{+1.5}_{-0.7}$          &   0.95     &   0.23    & B2012        \\
J111850G1      & 25.1    & 19.97$\pm$0.2         & $116 \pm 10$         & $175 \pm 80$            &  30 $\pm$ 5.0        &   7.0$^\dagger$       &  4.1 $^{+10.0}_{-1.1}$         &   0.63     &   1.17    & B2012        \\
J225036G1      & 53.9    & 19.54$\pm$0.2         & $240 \pm 10$         & $225 \pm 50$            &  30 $\pm$ 5.0        &   8.0$^\dagger$       &  4.2 $^{+7.3}_{-2.2}$          &   0.40     &   1.06    & B2012        \\
J1659$+$3735   & 58.0    & 18.89$\pm$0.15        & $140 \pm 10$         & 40--80                  &  40 $\pm$ 5.0        &   4.6--15             &  1.6--4.2                 &   0.12--0.27 & 0.1--0.9& K2014\\
\hline
\end{tabular}\\
{
(1) Galaxy name;
(2) Impact parameter (kpc);
(3) Gas column density at the impact parameter (cm$^{-2}$);
(4) Maximum rotational velocity of the galaxy (\kms ); 
(5) Wind velocity (\kms );
(6) Cone opening angle (degrees)
(7) Star Formation Rate (\mpy);
(8) Ejected mass rate \textbf{for one cone} (\mpy);
(9) Ejection velocity \textbf{divided by} escape velocity;
(10) \textbf{Mass} loading factor: ejected mass rate \textbf{divided by} star formation rate 
(for both cones);
(11) References: B2012: \citet{bouche_12}, K2014: \citet{kacprzak_14}.\\
$^\dagger$ SED-derived SFRs.\\
}
\end{table*}

The ratio $V_{\rm out}/V_{\rm esc}<1$ shows that the ejected gas does not escape from the galaxy halo and should \textbf{therefore} fall back into the galaxy.
\textbf{This gas contributes to the regulation of star formation in the galaxies. }
Other cases may be possible: for a galaxy with a low inclination (nearly face on) and a low impact parameter, the quasar LOS can easily track outflowing and inflowing materials.

\section{Conclusions and discussions}
\label{section:conclusions}
\begin{figure*}[h!]
  \centering
  \includegraphics[width=14cm]{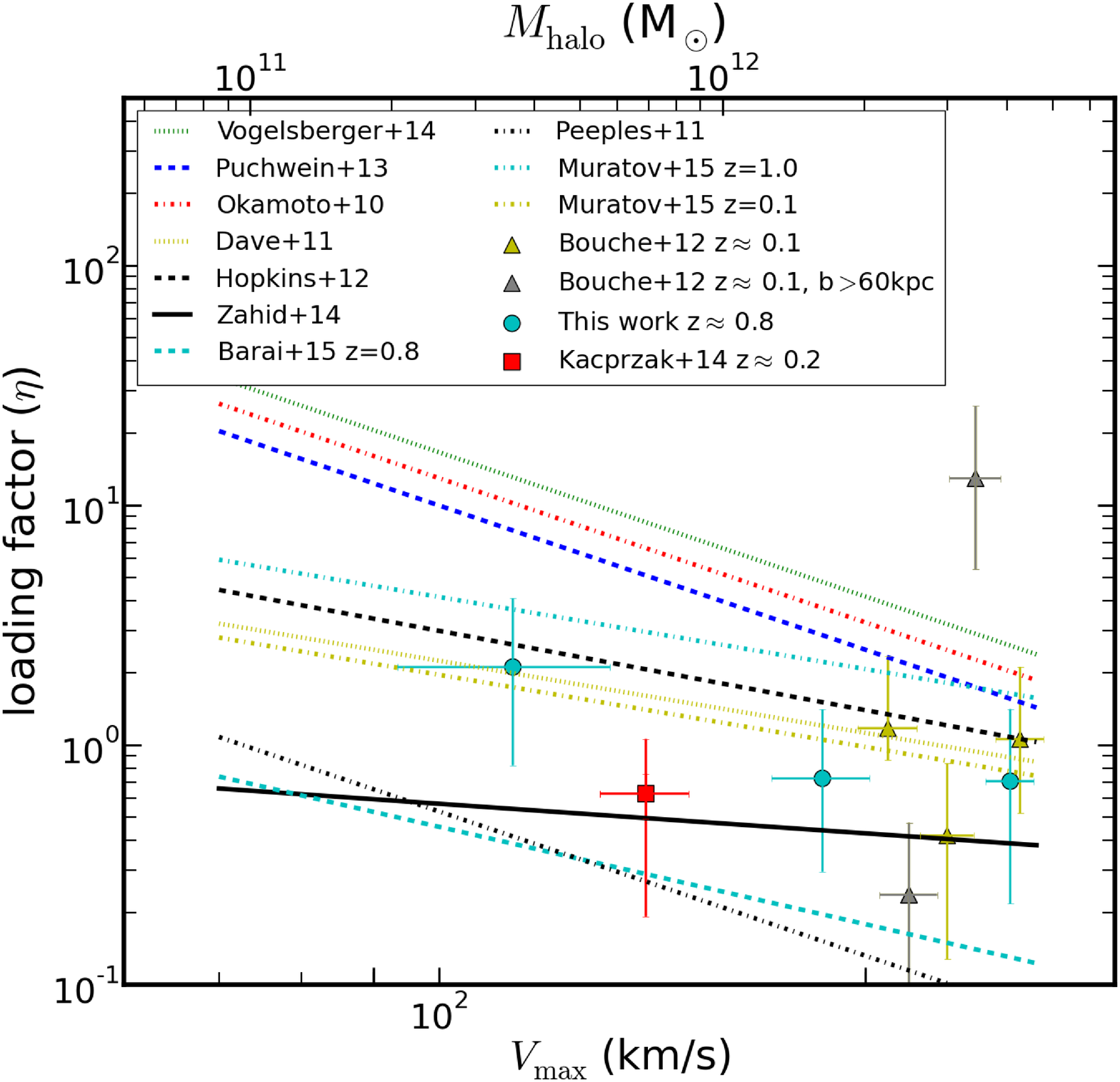}
  \caption{Comparison of predicted mass loading factors from theoretical/empirical models (curves) \textbf{with values derived from observations} (dots and triangles) as a function of the maximum rotational velocity.
 The results from this work are represented by the blue circles.
 The red circle shows the mass loading factor for a $z\sim0.2$ galaxy \citep{kacprzak_14}.
 The triangles show the results for $z\sim0.2$ galaxies from \citet{bouche_12}.
 The gray triangles show the galaxies with quasars located at $>$60kpc where the mass loading factor is less reliable due to the large travel time 
 needed for the outflow to cross the quasar LOS (several 100 Myr) compared to the short time scale of the \Ha\ derived SFR ($\sim 10$Myr).
 The upper halo mass axis is scaled on $V_{\rm max}$ at redshift 0.8 from \citet{mo_02}.
  }
  \label{fig:eta_vs_vmax}
\end{figure*}

In this paper, we studied gas flows around star-forming galaxies using the SIMPLE sample of galaxy$-$quasar pairs (paper~I).
The galaxies in this sample are located within $\leq$3\arcsec\ ($\leq$20~kpc) of the background quasar sightlines due to the selection of absorption with rest equivalent width $\geq$~2~\AA.
Thanks to the SINFONI IFU on the VLT and the new algorithm GalPaK$^{3D}$, we were able to recover the intrinsic morphological 
and kinematic properties of the galaxies from their \Ha\ emission (Fig.~1). 
The galaxies in our sample can be classified as \textbf{wind-pairs }
 or \textbf{inflow-pairs} 
according to the apparent location of the quasar
with respect to the \textbf{galaxy} major-axis (Fig.~\ref{fig:alpha}). 

With this classification, we focused the analysis on the properties of galactic winds for the sub-sample of  four suitable \textbf{wind-pairs}, 
\textbf{although one galaxy has} a SNR too low for a robust morphological (inclination) \textbf{measurement}.
The wind properties are constrained from the high-resolution UVES spectra of the minor-axis quasars. 
We show that a simple cone model for galactic winds (\ref{section:cone}) can reproduce the morphology of the UVES \MgI\ absorption profiles.
The wind properties can be summarized as follows:
\begin{itemize}
\item \textbf{Like other recent works \citep[e.g.][]{rubin_14, chisholm_14}}, outflow velocities are smaller than the escape velocities, so the gas traced by low-ionization lines does not escape the galaxy halo.
\item At the lowest impact parameter ($b\sim6$~kpc), one quasar-galaxy pair (J2357$-$2736) \textbf{has an absorption profile} consistent with an accelerated wind (\ref{section:accelerated}).
\item Loading factors $\eta$ vary between $\sim  0.65$ for the two massive galaxies and $\sim 2$ for the galaxy with the lowest mass (Figure~\ref{fig:eta_vs_vmax}).
Our results indicate that the mass loading factors tend to be higher for smaller galaxies, in agreement with theoretical expectations  \citep[e.g.][]{murray_11,hopkins_12}.
\end{itemize}

\textbf{
Figure~\ref{fig:eta_vs_vmax} also includes observational constraints on the mass loading factor from 
the $z\sim0.1$ survey of \citet[][]{bouche_12} (triangles) and $z\sim0.2$ pair of \citet{kacprzak_14} (square).
For  the \citet{bouche_12} sample, we used SED-derived SFRs (Table 4)  given their larger  impact parameters and longer travel times ($b/V_{\rm out}>$ 100s of Myr) compared to the time scale for \Ha-derived SFRs (few Myr).
The SED derived SFRs are computed from the UV-to-IR photometry (using Galex$+$SDSS$+$Wise surveys) with the Code Investigating GALaxy Emission (CIGALE) software \citep{nolls_09}.
The pairs with the largest impact parameters ($>60$ kpc) are shown in grey, since these mass loading factors can suffer strong biases due to even larger travel times ($\sim 300$ Myr).  
Other measurements at higher redshift from stacked spectra of star-forming galaxies \citep{weiner_09, newman_13} indicate an average mass loading factor of $\sim2$ \citep{newman_13}.
}

The different lines in Figure~\ref{fig:eta_vs_vmax} represent various theoretical \citep[][]{okamoto_10, dave_11b, hopkins_12, puchwein_springel_13, vogelsberger_14, muratov_15, barai_14} and 
empirical models \citep[][]{zahid_13b, peeples_shankar_11}~\footnote{The parameters of some of these models are listed in Table 1 of \citet{zahid_13b}. We also took the values of the corrected version of \citet{vogelsberger_14}}. 
In comparing observations and models, 
it is important to bear
in mind that some are fiducial scaling relations in the sub-grid physics \citep[e.g.][]{ dave_11b, puchwein_springel_13, vogelsberger_14}, while others are from more complex numerical approaches
\citep[e.g.][]{hopkins_12,hopkins_13,muratov_15, barai_14}~\footnote{Note that the outflow rate from \citet{barai_14} only includes gas particles with velocities greater than the escape velocity.} 
\textbf{, and thus are more directly comparable with observations.}

Currently our data does not allow us to discriminate between energy and momentum driven winds, but 
thanks to ongoing work at redshift $z\sim0.2$ with Keck/LRIS (Bouch\'e et al., in prep.; Martin et al., in prep.) 
and to the new generation IFU Multi Unit Spectroscopic Explorer (MUSE) on \textbf{the} VLT \citep{bacon_10, bacon_15}, we will be able to significantly increase the sample size and put tight constraints on the wind scaling relations.

\acknowledgments
\textbf{We would like to thank the referee for his/her thorough read of the manuscript and for the useful suggestions and comments.}
This work was partly supported by a Marie Curie International Career Integration Grant (PCIG11-GA-2012-321702).
This work is based on observations taken at ESO/VLT in Paranal and partially from the data archive of the NASA/ESA {\it HST}, which is
operated by the Association of Universities for Research in Astronomy, Inc.
We would like to thank the ESO staff.
NB acknowledges support from a Carreer Integration Grant (CIG) (PCIG11-GA-2012-321702) within the 7th European Community Framework Program.
MTM thanks the Australian Research Council for \textsl{Discovery Project} grant DP130100568 which supported this work.
CP thanks the ‘Agence Nationale de la Recherche’ for support (contract ANR-08-BLAN-0316-01).
This research made use of Astropy, a community-developed core 
PYTHON package for astronomy (Astropy Collaboration et al. 2013), NumPy and SciPy (Oliphant 2007), Matplotlib (Hunter 2007), IPython (Perez\& Granger 2007) and of
 NASA’s Astrophysics Data System Bibliographic Services.

\pagebreak
\appendix
\section{Understanding the geometric wind model}
\setcounter{figure}{0} \renewcommand{\thefigure}{A.\arabic{figure}} 

\begin{figure*}[h!]
  \centering
  \includegraphics[width=15cm]{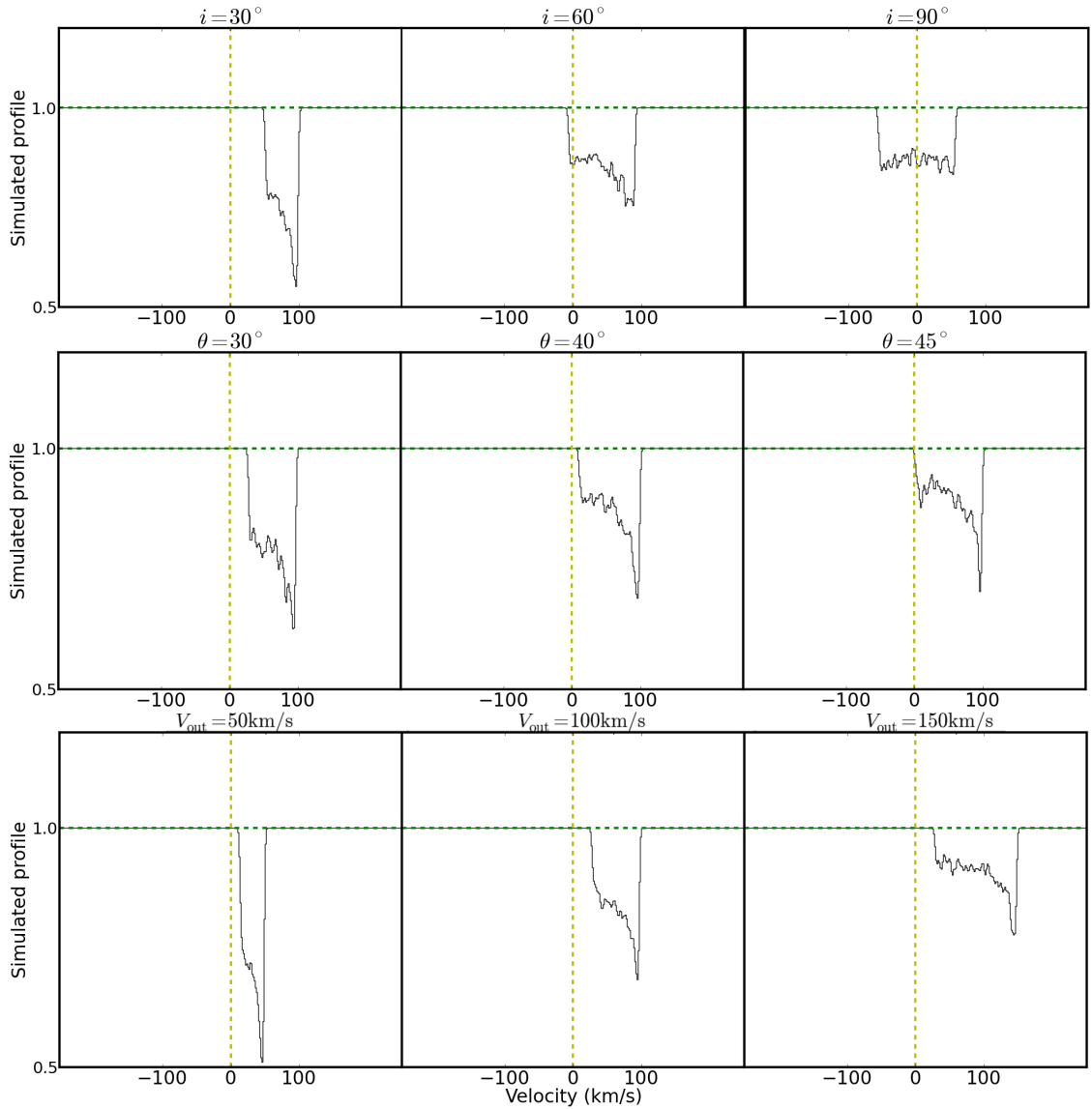}
  \caption{Examples of simulated \textbf{absorption} profiles with different galaxy inclinations ($i$), opening angle ($\theta$) and wind velocities ($V_{\rm out}$):
  \textbf{while each of the simulated profiles has the same number of particles, the apparent depth decreases as each parameter increases due to larger velocity projections for $i$ and $\theta$, 
  and larger range of velocities for $V_{\rm out}$.}
  \textit{top row}: \textbf{absorption profiles for galaxies inclined at} 30, 60 and 90 degrees with $V_{\rm out}$= 100 \kms and $\theta$= 30$^{\circ}$.
  The noise effect is due to \textbf{the} Monte Carlo distribution of particles.
  \textit{middle row}: \textbf{absorption profiles for wind cones with} opening angles of 30, 40 and 45 degrees with $V_{\rm out}$= 100 \kms and $i$=45$^{\circ}$.
  \textit{bottom row}: \textbf{absorption profiles with} wind velocities of 50, 100 and 150 \kms with $i$= 45 degrees and $\theta$= 30$^{\circ}$.
  Each simulated profile has the same amount of particles but show a larger velocity range due to the increasing gas speed, hence the varying apparent depths.
  }
  \label{fig:models}
\end{figure*}

In this appendix we \textbf{demonstrate how varying different parameters within the cone model impacts the simulated absorption profile.}
For the cone model \textbf{(Section~\ref{section:cone})}, we change three parameters in order to investigate the behavior of the simulated profile: for different \textbf{galaxy} inclinations ($i$), 
\textbf{cone} opening angles ($\theta$), and wind outflow velocities ($V_{\rm out}$)(Figure~\ref{fig:models}).
\textbf{The general trends are as follows. The different inclinations and cone opening angles change the left portion of the simulated absorption profile, while the wind velocity extends the right portion. 
Note that, except for the case of 90$^{\circ}$ galaxy inclination, all the simulated absorption profiles in Figure~\ref{fig:models} present an outward asymmetry.}

\begin{figure*}[h!]
  \centering
  \includegraphics[width=16cm]{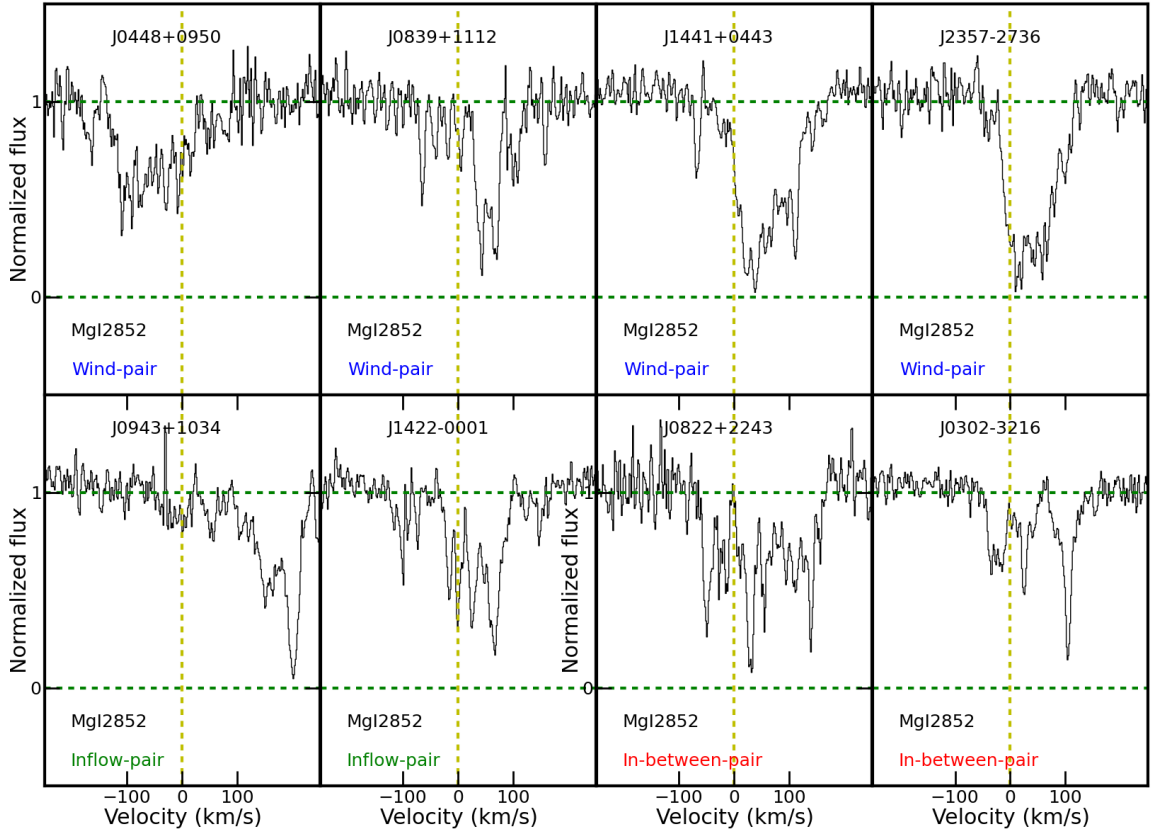}
  \caption{Here we present the UVES \MgI\ \mc\ absorption lines for 8 out of 10 galaxies. We do not have these absorption lines for J0147+1258 and J0226$-$2857 as they fall in the gap of UVES data.
  The top row corresponds to UVES \MgI\ centered data of the four wind-pairs.
  The bottom row presents the UVES \MgI\ centered data of the two inflow-pairs and the two ambiguous cases from left to right.
  }
  \label{fig:mgi}
\end{figure*}

\textbf{We also present in Figure~\ref{fig:mgi} the UVES \MgI\ \mc\ absorption lines of our SIMPLE sample galaxies.}

\begin{thebibliography}{95}
\expandafter\ifx\csname natexlab\endcsname\relax\def\natexlab#1{#1}\fi

\bibitem[{{Abuter} {et~al.}(2006){Abuter}, {Schreiber}, {Eisenhauer}, {Ott},
  {Horrobin}, \& {Gillesen}}]{AbuterR_06a}
{Abuter}, R., {Schreiber}, J., {Eisenhauer}, F., {Ott}, T., {Horrobin}, M., \&
  {Gillesen}, S. 2006, New Astronomy Review, 50, 398
  
\bibitem[Amram et al.(2013)]{amram_13} Amram, P., 
L{\'o}pez-Sanjuan, C., Epinat, B., et al.\ 2013, IAU Symposium, 295, 86

\bibitem[{{Bacon} {et~al.}(2010){Bacon}, {Accardo}, {Adjali}, {Anwand},
  {Bauer}, {Biswas}, {Blaizot}, {Boudon}, {Brau-Nogue}, {Brinchmann},
  {Caillier}, {Capoani}, {Carollo}, {Contini}, {Couderc}, {Daguis{\'e}},
  {Deiries}, {Delabre}, {Dreizler}, {Dubois}, {Dupieux}, {Dupuy}, {Emsellem},
  {Fechner}, {Fleischmann}, {Fran{\c c}ois}, {Gallou}, {Gharsa}, {Glindemann},
  {Gojak}, {Guiderdoni}, {Hansali}, {Hahn}, {Jarno}, {Kelz}, {Koehler},
  {Kosmalski}, {Laurent}, {Le Floch}, {Lilly}, {Lizon}, {Loupias}, {Manescau},
  {Monstein}, {Nicklas}, {Olaya}, {Pares}, {Pasquini}, {P{\'e}contal-Rousset},
  {Pell{\'o}}, {Petit}, {Popow}, {Reiss}, {Remillieux}, {Renault}, {Roth},
  {Rupprecht}, {Serre}, {Schaye}, {Soucail}, {Steinmetz}, {Streicher}, {Stuik},
  {Valentin}, {Vernet}, {Weilbacher}, {Wisotzki}, \& {Yerle}}]{bacon_10}
{Bacon}, R. {et~al.} 2010, in Society of Photo-Optical Instrumentation
  Engineers (SPIE) Conference Series, Vol. 7735, Society of Photo-Optical
  Instrumentation Engineers (SPIE) Conference Series, 8

\bibitem[Bacon et al.(2014)]{bacon_15} Bacon, R., Brinchmann, 
J., Richard, J., et al.\ 2014, arXiv:1411.7667
  
\bibitem[Barai et al.(2015)]{barai_14} Barai, P., Monaco, P., 
Murante, G., Ragagnin, A., \& Viel, M.\ 2015, \mnras, 447, 266

\bibitem[{{Bardeen} {et~al.}(1983){Bardeen}, {Steinhardt}, \&
  {Turner}}]{bardeen_83}
{Bardeen}, J.~M., {Steinhardt}, P.~J., \& {Turner}, M.~S. 1983, \prd, 28, 679

\bibitem[{{Behroozi} {et~al.}(2010){Behroozi}, {Conroy}, \&
  {Wechsler}}]{behroozi_10}
{Behroozi}, P.~S., {Conroy}, C., \& {Wechsler}, R.~H. 2010, \apj, 717, 379

\bibitem[{{Benson} {et~al.}(2003){Benson}, {Bower}, {Frenk}, {Lacey}, {Baugh},
  \& {Cole}}]{benson_03}
{Benson}, A.~J., {Bower}, R.~G., {Frenk}, C.~S., {Lacey}, C.~G., {Baugh},
  C.~M., \& {Cole}, S. 2003, \apj, 599, 38

\bibitem[{{Bergeron}(1988)}]{bergeron_88}
{Bergeron}, J. 1988, in IAU Symposium, Vol. 130, Large Scale Structures of the
  Universe, ed. J.~{Audouze}, M.-C. {Pelletan}, A.~{Szalay}, Y.~B.
  {Zel'dovich}, \& P.~J.~E. {Peebles}, 343

\bibitem[{{Bergeron} \& {Boisse}(1991)}]{bergeron_91}
{Bergeron}, J., \& {Boisse}, P. 1991, Advances in Space Research, 11, 241

\bibitem[{{Bernardi} {et~al.}(2010){Bernardi}, {Shankar}, {Hyde}, {Mei},
  {Marulli}, \& {Sheth}}]{bernardi_10}
{Bernardi}, M., {Shankar}, F., {Hyde}, J.~B., {Mei}, S., {Marulli}, F., \&
  {Sheth}, R.~K. 2010, \mnras, 404, 2087
  
\bibitem[Bonnet et al.(2004)]{bonnet_04} Bonnet, H., Abuter, R., 
Baker, A., et al.\ 2004, The Messenger, 117, 17

\bibitem[{{Bordoloi} {et~al.}(2014){Bordoloi}, {Lilly}, {Hardmeier}, {Contini},
  {Kneib}, {Le Fevre}, {Mainieri}, {Renzini}, {Scodeggio}, {Zamorani},
  {Bardelli}, {Bolzonella}, {Bongiorno}, {Caputi}, {Carollo}, {Cucciati}, {de
  la Torre}, {de Ravel}, {Garilli}, {Iovino}, {Kampczyk}, {Kova{\v c}},
  {Knobel}, {Lamareille}, {Le Borgne}, {Le Brun}, {Maier}, {Mignoli}, {Oesch},
  {Pello}, {Peng}, {Perez Montero}, {Presotto}, {Silverman}, {Tanaka}, {Tasca},
  {Tresse}, {Vergani}, {Zucca}, {Cappi}, {Cimatti}, {Coppa}, {Franzetti},
  {Koekemoer}, {Moresco}, {Nair}, \& {Pozzetti}}]{bordoloi_14}
{Bordoloi}, R. {et~al.} 2014, \apj, 794, 130

\bibitem[Bohlin et al.(1978)]{Bohlin_78} Bohlin, R.~C., Savage, 
B.~D., \& Drake, J.~F.\ 1978, \apj, 224, 132 

\bibitem[{{Bordoloi} {et~al.}(2011){Bordoloi}, {Lilly}, {Knobel}, {Bolzonella},
  {Kampczyk}, {Carollo}, {Iovino}, {Zucca}, {Contini}, {Kneib}, {Le Fevre},
  {Mainieri}, {Renzini}, {Scodeggio}, {Zamorani}, {Balestra}, {Bardelli},
  {Bongiorno}, {Caputi}, {Cucciati}, {de la Torre}, {de Ravel}, {Garilli},
  {Kova{\v c}}, {Lamareille}, {Le Borgne}, {Le Brun}, {Maier}, {Mignoli},
  {Pello}, {Peng}, {Perez Montero}, {Presotto}, {Scarlata}, {Silverman},
  {Tanaka}, {Tasca}, {Tresse}, {Vergani}, {Barnes}, {Cappi}, {Cimatti},
  {Coppa}, {Diener}, {Franzetti}, {Koekemoer}, {L{\'o}pez-Sanjuan},
  {McCracken}, {Moresco}, {Nair}, {Oesch}, {Pozzetti}, \&
  {Welikala}}]{bordoloi_11}
---. 2011, \apj, 743, 10

\bibitem[{{Bouch{\'e}} {et~al.}(2012){Bouch{\'e}}, {Hohensee}, {Vargas},
  {Kacprzak}, {Martin}, {Cooke}, \& {Churchill}}]{bouche_12}
{Bouch{\'e}}, N., {Hohensee}, W., {Vargas}, R., {Kacprzak}, G.~G., {Martin},
  C.~L., {Cooke}, J., \& {Churchill}, C.~W. 2012, \mnras, 426, 801

\bibitem[{{Bouch{\'e}} {et~al.}(2013){Bouch{\'e}}, {Murphy}, {Kacprzak},
  {P{\'e}roux}, {Contini}, {Martin}, \& {Dessauges-Zavadsky}}]{bouche_13}
{Bouch{\'e}}, N., {Murphy}, M.~T., {Kacprzak}, G.~G., {P{\'e}roux}, C.,
  {Contini}, T., {Martin}, C.~L., \& {Dessauges-Zavadsky}, M. 2013, Science,
  341, 50

\bibitem[{{Bouch{\'e}} {et~al.}(2006){Bouch{\'e}}, {Murphy}, {P{\'e}roux},
  {Csabai}, \& {Wild}}]{bouche_06}
{Bouch{\'e}}, N., {Murphy}, M.~T., {P{\'e}roux}, C., {Csabai}, I., \& {Wild},
  V. 2006, \mnras, 371, 495

\bibitem[{{Bouch{\'e}} {et~al.}(2007){Bouch{\'e}}, {Murphy}, {P{\'e}roux},
  {Davies}, {Eisenhauer}, {F{\"o}rster Schreiber}, \& {Tacconi}}]{bouche_07}
{Bouch{\'e}}, N., {Murphy}, M.~T., {P{\'e}roux}, C., {Davies}, R.,
  {Eisenhauer}, F., {F{\"o}rster Schreiber}, N.~M., \& {Tacconi}, L. 2007,
  \apjl, 669, L5

\bibitem[Bouch{\'e} et al.(2015)]{bouche_15} Bouch{\'e}, N., 
Carfantan, H., Schroetter, I., Michel-Dansac, L., 
\& Contini, T.\ 2015, arXiv:1501.06586
  
\bibitem[{{Bower} {et~al.}(2006){Bower}, {Benson}, {Malbon}, {Helly}, {Frenk},
  {Baugh}, {Cole}, \& {Lacey}}]{bower_06}
{Bower}, R.~G., {Benson}, A.~J., {Malbon}, R., {Helly}, J.~C., {Frenk}, C.~S.,
  {Baugh}, C.~M., {Cole}, S., \& {Lacey}, C.~G. 2006, \mnras, 370, 645

\bibitem[Ceverino et al.(2010)]{ceverino_10} Ceverino, D., Dekel, 
A., \& Bournaud, F.\ 2010, \mnras, 404, 2151

\bibitem[{{Chabrier}(2003)}]{chabrier_03}
{Chabrier}, G. 2003, \pasp, 115, 763

\bibitem[{{Chen}(2012)}]{chenhw_12}
{Chen}, H.-W. 2012, \mnras, 427, 1238

\bibitem[Chisholm et al.(2014)]{chisholm_14} Chisholm, J., 
Tremonti, C.~A., Leitherer, C., et al.\ 2014, arXiv:1412.2139 

\bibitem[Contini et 
al.(2012)]{contini_12} Contini, T., Garilli, B., Le F{\`e}vre, O., et al.\ 2012, \aap, 539, AA91

\bibitem[{{Cresci} {et~al.}(2009){Cresci}, {Hicks}, {Genzel}, {Schreiber},
  {Davies}, {Bouch{\'e}}, {Buschkamp}, {Genel}, {Shapiro}, {Tacconi},
  {Sommer-Larsen}, {Burkert}, {Eisenhauer}, {Gerhard}, {Lutz}, {Naab},
  {Sternberg}, {Cimatti}, {Daddi}, {Erb}, {Kurk}, {Lilly}, {Renzini},
  {Shapley}, {Steidel}, \& {Caputi}}]{cresci_09}
{Cresci}, G. {et~al.} 2009, \apj, 697, 115

\bibitem[{{Croton} {et~al.}(2006){Croton}, {Springel}, {White}, {De Lucia},
  {Frenk}, {Gao}, {Jenkins}, {Kauffmann}, {Navarro}, \& {Yoshida}}]{croton_06}
{Croton}, D.~J. {et~al.} 2006, \mnras, 365, 11

\bibitem[{{Dav{\'e}} {et~al.}(2011){Dav{\'e}}, {Oppenheimer}, \&
  {Finlator}}]{dave_11b}
{Dav{\'e}}, R., {Oppenheimer}, B.~D., \& {Finlator}, K. 2011, \mnras, 415, 11

\bibitem[{{Davies}(2007)}]{DaviesR_06a}
{Davies}, R. 2007, \mnras, 375, 1099

\bibitem[Dekker et al.(2000)]{dekker_00} Dekker, H., D'Odorico, 
S., Kaufer, A., Delabre, B., \& Kotzlowski, H.\ 2000, \procspie, 4008, 534 

\bibitem[{{Dekel} \& {Silk}(1986)}]{dekel_86}
{Dekel}, A., \& {Silk}, J. 1986, \apj, 303, 39

\bibitem[Dekel et al.(2009)]{dekel_09} Dekel, A., Sari, R., 
\& Ceverino, D.\ 2009, \apj, 703, 785

\bibitem[{{Dekel} {et~al.}(2013){Dekel}, {Zolotov}, {Tweed}, {Cacciato},
  {Ceverino}, \& {Primack}}]{dekel_13}
{Dekel}, A., {Zolotov}, A., {Tweed}, D., {Cacciato}, M., {Ceverino}, D., \&
  {Primack}, J.~R. 2013, \mnras, 435, 999

\bibitem[{{Dubois} \& {Teyssier}(2008)}]{dubois_08}
{Dubois}, Y., \& {Teyssier}, R. 2008, \aap, 477, 79

\bibitem[Elmegreen et al.(2007)]{elmegreen_07} Elmegreen, D.~M., 
Elmegreen, B.~G., Ravindranath, S., \& Coe, D.~A.\ 2007, \apj, 658, 763

\bibitem[{{Epinat} {et~al.}(2009){Epinat}, {Contini}, {Le F{\`e}vre},
  {Vergani}, {Garilli}, {Amram}, {Queyrel}, {Tasca}, \& {Tresse}}]{epinat_09}
{Epinat}, B. {et~al.} 2009, \aap, 504, 789

\bibitem[{{Epinat} {et~al.}(2012){Epinat}, {Tasca}, {Amram}, {Contini}, {Le
  F{\`e}vre}, {Queyrel}, {Vergani}, {Garilli}, {Kissler-Patig}, {Moultaka},
  {Paioro}, {Tresse}, {Bournaud}, {L{\'o}pez-Sanjuan}, \& {Perret}}]{epinat_12}
---. 2012, \aap, 539, A92

\bibitem[F{\"o}rster Schreiber et al.(2006)]{forsterschreiber_06} 
F{\"o}rster Schreiber, N.~M., Genzel, R., Eisenhauer, F., et al.\ 2006, The 
Messenger, 125, 11

\bibitem[{{F{\"o}rster Schreiber} {et~al.}(2009{\natexlab{a}}){F{\"o}rster
  Schreiber}, {Genzel}, {Bouch{\'e}}, {Cresci}, {Davies}, {Buschkamp},
  {Shapiro}, {Tacconi}, {Hicks}, {Genel}, \& {Shapley}}]{ForsterSchreiberN_09a}
{F{\"o}rster Schreiber}, N.~M. {et~al.} 2009{\natexlab{a}}, \apj, 706, 1364

\bibitem[{{F{\"o}rster Schreiber} {et~al.}(2009{\natexlab{b}}){F{\"o}rster
  Schreiber}, {Genzel}, {Bouch{\'e}}, {Cresci}, {Davies}, {Buschkamp},
  {Shapiro}, {Tacconi}, {Hicks}, {Genel}, {Shapley}, {Erb}, {Steidel}, {Lutz},
  {Eisenhauer}, {Gillessen}, {Sternberg}, {Renzini}, {Cimatti}, {Daddi},
  {Kurk}, {Lilly}, {Kong}, {Lehnert}, {Nesvadba}, {Verma}, {McCracken},
  {Arimoto}, {Mignoli}, \& {Onodera}}]{forsterschreiber_09}
---. 2009{\natexlab{b}}, \apj, 706, 1364

\bibitem[{{F{\"o}rster Schreiber} {et~al.}(2011){F{\"o}rster Schreiber},
  {Shapley}, {Genzel}, {Bouch{\'e}}, {Cresci}, {Davies}, {Erb}, {Genel},
  {Lutz}, {Newman}, {Shapiro}, {Steidel}, {Sternberg}, \&
  {Tacconi}}]{forsterschreiber_11}
---. 2011, \apj, 739, 45

\bibitem[Genel et al.(2012)]{genel_12} Genel, S., Naab, T., 
Genzel, R., et al.\ 2012, \apj, 745, 11

\bibitem[Genel et al.(2014)]{genel_14} Genel, S., Vogelsberger, 
M., Springel, V., et al.\ 2014, \mnras, 445, 175

\bibitem[{{Genzel} {et~al.}(2008){Genzel}, {Burkert}, {Bouch{\'e}}, {Cresci},
  {F{\"o}rster Schreiber}, {Shapley}, {Shapiro}, {Tacconi}, {Buschkamp},
  {Cimatti}, {Daddi}, {Davies}, {Eisenhauer}, {Erb}, {Genel}, {Gerhard},
  {Hicks}, {Lutz}, {Naab}, {Ott}, {Rabien}, {Renzini}, {Steidel}, {Sternberg},
  \& {Lilly}}]{genzel_08}
{Genzel}, R. {et~al.} 2008, \apj, 687, 59

\bibitem[{{Genzel} {et~al.}(2011){Genzel}, {Newman}, {Jones}, {F{\"o}rster
  Schreiber}, {Shapiro}, {Genel}, {Lilly}, {Renzini}, {Tacconi}, {Bouch{\'e}},
  {Burkert}, {Cresci}, {Buschkamp}, {Carollo}, {Ceverino}, {Davies}, {Dekel},
  {Eisenhauer}, {Hicks}, {Kurk}, {Lutz}, {Mancini}, {Naab}, {Peng},
  {Sternberg}, {Vergani}, \& {Zamorani}}]{genzel_11}
---. 2011, \apj, 733, 101

\bibitem[{{Genzel} {et~al.}(2006){Genzel}, {Tacconi}, {Eisenhauer},
  {F{\"o}rster Schreiber}, {Cimatti}, {Daddi}, {Bouch{\'e}}, {Davies},
  {Lehnert}, {Lutz}, {Nesvadba}, {Verma}, {Abuter}, {Shapiro}, {Sternberg},
  {Renzini}, {Kong}, {Arimoto}, \& {Mignoli}}]{genzel_06}
---. 2006, \nat, 442, 786

\bibitem[{{Guth} \& {Pi}(1982)}]{guth_82}
{Guth}, A.~H., \& {Pi}, S.-Y. 1982, Physical Review Letters, 49, 1110

\bibitem[{{Heckman} {et~al.}(1990){Heckman}, {Armus}, \& {Miley}}]{heckman_90}
{Heckman}, T.~M., {Armus}, L., \& {Miley}, G.~K. 1990, \apjs, 74, 833

\bibitem[{{Heckman} {et~al.}(2000){Heckman}, {Lehnert}, {Strickland}, \&
  {Armus}}]{heckman_00}
{Heckman}, T.~M., {Lehnert}, M.~D., {Strickland}, D.~K., \& {Armus}, L. 2000,
  \apjs, 129, 493

\bibitem[{{Hopkins} {et~al.}(2012){Hopkins}, {Quataert}, \&
  {Murray}}]{hopkins_12}
{Hopkins}, P.~F., {Quataert}, E., \& {Murray}, N. 2012, \mnras, 421, 3522

\bibitem[Hopkins et al.(2013)]{hopkins_13} Hopkins, P.~F., Kere{\v 
s}, D., Murray, N., et al.\ 2013, \mnras, 433, 78

\bibitem[{{Jenkins}(2009)}]{jenkins_09}
{Jenkins}, E.~B. 2009, \apj, 700, 1299

\bibitem[Kacprzak et al.(2010)]{Kacprzak_10a} Kacprzak, G.~G., 
Churchill, C.~W., Ceverino, D., et al.\ 2010, \apj, 711, 533

\bibitem[{{Kacprzak} {et~al.}(2011{\natexlab{a}}){Kacprzak}, {Churchill},
  {Barton}, \& {Cooke}}]{kacprzak_11}
{Kacprzak}, G.~G., {Churchill}, C.~W., {Barton}, E.~J., \& {Cooke}, J.
  2011{\natexlab{a}}, \apj, 733, 105

\bibitem[{{Kacprzak} {et~al.}(2011{\natexlab{b}}){Kacprzak}, {Churchill},
  {Evans}, {Murphy}, \& {Steidel}}]{kacprzak_11b}
{Kacprzak}, G.~G., {Churchill}, C.~W., {Evans}, J.~L., {Murphy}, M.~T., \&
  {Steidel}, C.~C. 2011{\natexlab{b}}, \mnras, 416, 3118


\bibitem[Kacprzak et al.(2014)]{kacprzak_14} Kacprzak, G.~G., 
Martin, C.~L., Bouch{\'e}, N., et al.\ 2014, \apjl, 792, LL12 

\bibitem[{{Kennicutt}(1998)}]{kennicutt_98}
{Kennicutt}, Jr., R.~C. 1998, \apj, 498, 541

\bibitem[Kere{\v s} et al.(2005)]{keres_05} Kere{\v s}, D., 
Katz, N., Weinberg, D.~H., \& Dav{\'e}, R.\ 2005, \mnras, 363, 2

\bibitem[{{Lan} {et~al.}(2014){Lan}, {M{\'e}nard}, \& {Zhu}}]{lan_menard_14}
{Lan}, T.-W., {M{\'e}nard}, B., \& {Zhu}, G. 2014, \apj, 795, 31

\bibitem[{{Law} {et~al.}(2007){Law}, {Steidel}, {Erb}, {Larkin}, {Pettini},
  {Shapley}, \& {Wright}}]{law_07}
{Law}, D.~R., {Steidel}, C.~C., {Erb}, D.~K., {Larkin}, J.~E., {Pettini}, M.,
  {Shapley}, A.~E., \& {Wright}, S.~A. 2007, \apj, 669, 929

\bibitem[{{Law} {et~al.}(2009){Law}, {Steidel}, {Erb}, {Larkin}, {Pettini},
  {Shapley}, \& {Wright}}]{law_09}
---. 2009, \apj, 697, 2057

\bibitem[{{Leauthaud} {et~al.}(2012){Leauthaud}, {George}, {Behroozi}, {Bundy},
  {Tinker}, {Wechsler}, {Conroy}, {Finoguenov}, \& {Tanaka}}]{leauthaud_12}
{Leauthaud}, A. {et~al.} 2012, \apj, 746, 95

\bibitem[{{Leauthaud} {et~al.}(2011){Leauthaud}, {Tinker}, {Behroozi}, {Busha},
  \& {Wechsler}}]{leauthaud_11}
{Leauthaud}, A., {Tinker}, J., {Behroozi}, P.~S., {Busha}, M.~T., \&
  {Wechsler}, R.~H. 2011, \apj, 738, 45

\bibitem[{{Lehnert} \& {Heckman}(1996)}]{heckman_96}
{Lehnert}, M.~D., \& {Heckman}, T.~M. 1996, \apj, 472, 546

\bibitem[{{Lehnert} {et~al.}(1999){Lehnert}, {van Breugel}, {Heckman}, \&
  {Miley}}]{lehnert_99}
{Lehnert}, M.~D., {van Breugel}, W.~J.~M., {Heckman}, T.~M., \& {Miley}, G.~K.
  1999, \apjs, 124, 11


\bibitem[{{Martin}(1998)}]{martin_98}
{Martin}, C.~L. 1998, \apj, 506, 222

\bibitem[{{Martin}(1999)}]{martin_99}
---. 1999, \apj, 513, 156

\bibitem[{{Martin}(2005)}]{martin_05}
---. 2005, \apj, 621, 227

\bibitem[{{Martin} {et~al.}(2002){Martin}, {Kobulnicky}, \&
  {Heckman}}]{martin_02}
{Martin}, C.~L., {Kobulnicky}, H.~A., \& {Heckman}, T.~M. 2002, \apj, 574, 663

\bibitem[{{Martin} {et~al.}(2012){Martin}, {Shapley}, {Coil}, {Kornei},
  {Bundy}, {Weiner}, {Noeske}, \& {Schiminovich}}]{martin_12}
{Martin}, C.~L., {Shapley}, A.~E., {Coil}, A.~L., {Kornei}, K.~A., {Bundy}, K.,
  {Weiner}, B.~J., {Noeske}, K.~G., \& {Schiminovich}, D. 2012, \apj, 760, 127

\bibitem[{{Martin} {et~al.}(2013){Martin}, {Shapley}, {Coil}, {Kornei},
  {Murray}, \& {Pancoast}}]{martin_13}
{Martin}, C.~L., {Shapley}, A.~E., {Coil}, A.~L., {Kornei}, K.~A., {Murray},
  N., \& {Pancoast}, A. 2013, \apj, 770, 41

\bibitem[{{M{\'e}nard} \& {Chelouche}(2009)}]{menard_09}
{M{\'e}nard}, B., \& {Chelouche}, D. 2009, \mnras, 393, 808

\bibitem[Mo \& White(2002)]{mo_02} Mo, H.~J., \& White, S.~D.~M.\ 2002, \mnras, 336, 112 

\bibitem[{{Moster} {et~al.}(2010){Moster}, {Somerville}, {Maulbetsch}, {van den
  Bosch}, {Macci{\`o}}, {Naab}, \& {Oser}}]{moster_10}
{Moster}, B.~P., {Somerville}, R.~S., {Maulbetsch}, C., {van den Bosch}, F.~C.,
  {Macci{\`o}}, A.~V., {Naab}, T., \& {Oser}, L. 2010, \apj, 710, 903

\bibitem[Muratov et al.(2015)]{muratov_15} Muratov, A.~L., Keres, 
D., Faucher-Giguere, C.-A., et al.\ 2015, arXiv:1501.03155 

\bibitem[{{Murray} {et~al.}(2011){Murray}, {M{\'e}nard}, \&
  {Thompson}}]{murray_11}
{Murray}, N., {M{\'e}nard}, B., \& {Thompson}, T.~A. 2011, \apj, 735, 66

\bibitem[{{Murray} {et~al.}(2005){Murray}, {Quataert}, \&
  {Thompson}}]{murray_05}
{Murray}, N., {Quataert}, E., \& {Thompson}, T.~A. 2005, \apj, 618, 569

\bibitem[{{Newman} {et~al.}(2012){Newman}, {Genzel}, {F{\"o}rster-Schreiber},
  {Shapiro Griffin}, {Mancini}, {Lilly}, {Renzini}, {Bouch{\'e}}, {Burkert},
  {Buschkamp}, {Carollo}, {Cresci}, {Davies}, {Eisenhauer}, {Genel}, {Hicks},
  {Kurk}, {Lutz}, {Naab}, {Peng}, {Sternberg}, {Tacconi}, {Vergani}, {Wuyts},
  \& {Zamorani}}]{newman_13}
{Newman}, S.~F. {et~al.} 2012, \apj, 761, 43

\bibitem[{{Noll} {et~al.}(2009){Noll}, {Burgarella}, {Giovannoli}, {Buat},
  {Marcillac}, \& {Mu{\~n}oz-Mateos}}]{nolls_09}
{Noll}, S., {Burgarella}, D., {Giovannoli}, E., {Buat}, V., {Marcillac}, D., \&
  {Mu{\~n}oz-Mateos}, J.~C. 2009, \aap, 507, 1793

\bibitem[{{Okamoto} {et~al.}(2010){Okamoto}, {Frenk}, {Jenkins}, \&
  {Theuns}}]{okamoto_10}
{Okamoto}, T., {Frenk}, C.~S., {Jenkins}, A., \& {Theuns}, T. 2010, \mnras,
  406, 208

\bibitem[{{Oppenheimer} \& {Dav{\'e}}(2006)}]{oppenheimer_06}
{Oppenheimer}, B.~D., \& {Dav{\'e}}, R. 2006, \mnras, 373, 1265

\bibitem[{{Oppenheimer} {et~al.}(2010){Oppenheimer}, {Dav{\'e}}, {Kere{\v s}},
  {Fardal}, {Katz}, {Kollmeier}, \& {Weinberg}}]{oppenheimer_10}
{Oppenheimer}, B.~D., {Dav{\'e}}, R., {Kere{\v s}}, D., {Fardal}, M., {Katz},
  N., {Kollmeier}, J.~A., \& {Weinberg}, D.~H. 2010, \mnras, 406, 2325

\bibitem[{{Papastergis} {et~al.}(2012){Papastergis}, {Cattaneo}, {Huang},
  {Giovanelli}, \& {Haynes}}]{papastergis_12}
{Papastergis}, E., {Cattaneo}, A., {Huang}, S., {Giovanelli}, R., \& {Haynes},
  M.~P. 2012, \apj, 759, 138

\bibitem[{{Peeples} \& {Shankar}(2011)}]{peeples_shankar_11}
{Peeples}, M.~S., \& {Shankar}, F. 2011, \mnras, 417, 2962

\bibitem[{{Peng} {et~al.}(2010){Peng}, {Ho}, {Impey}, \& {Rix}}]{peng_10}
{Peng}, C.~Y., {Ho}, L.~C., {Impey}, C.~D., \& {Rix}, H.-W. 2010, \aj, 139,
  2097

\bibitem[{{P{\'e}roux} {et~al.}(2013){P{\'e}roux}, {Bouch{\'e}}, {Kulkarni}, \&
  {York}}]{peroux_13}
{P{\'e}roux}, C., {Bouch{\'e}}, N., {Kulkarni}, V.~P., \& {York}, D.~G. 2013,
  \mnras, 436, 2650

\bibitem[Perret et al.(2012)]{perret_12} Perret, V., Epinat, B., 
Amram, P., et al.\ 2012, arXiv:1212.1356
  
\bibitem[{{Pettini} {et~al.}(2002){Pettini}, {Rix}, {Steidel}, {Adelberger},
  {Hunt}, \& {Shapley}}]{pettini_02}
{Pettini}, M., {Rix}, S.~A., {Steidel}, C.~C., {Adelberger}, K.~L., {Hunt},
  M.~P., \& {Shapley}, A.~E. 2002, \apj, 569, 742

\bibitem[{{Puchwein} \& {Springel}(2013)}]{puchwein_springel_13}
{Puchwein}, E., \& {Springel}, V. 2013, \mnras, 428, 2966

\bibitem[Queyrel et al.(2008)]{queyrel_08} Queyrel, J., Contini, 
T., Epinat, B., et al.\ 2008, SF2A-2008, 383

\bibitem[{{Rosdahl} {et~al.}(2013){Rosdahl}, {Blaizot}, {Aubert}, {Stranex}, \&
  {Teyssier}}]{rosdahl_13}
{Rosdahl}, J., {Blaizot}, J., {Aubert}, D., {Stranex}, T., \& {Teyssier}, R.
  2013, \mnras, 436, 2188
  
\bibitem[Ro{\v s}kar et al.(2014)]{roskar_13} Ro{\v s}kar, R., 
Teyssier, R., Agertz, O., Wetzstein, M., 
\& Moore, B.\ 2014, \mnras, 444, 2837 

\bibitem[{{Rubin} {et~al.}(2014){Rubin}, {Prochaska}, {Koo}, {Phillips},
  {Martin}, \& {Winstrom}}]{rubin_14}
{Rubin}, K.~H.~R., {Prochaska}, J.~X., {Koo}, D.~C., {Phillips}, A.~C.,
  {Martin}, C.~L., \& {Winstrom}, L.~O. 2014, \apj, 794, 156

\bibitem[{{Rubin} {et~al.}(2010){Rubin}, {Weiner}, {Koo}, {Martin},
  {Prochaska}, {Coil}, \& {Newman}}]{rubin_10}
{Rubin}, K.~H.~R., {Weiner}, B.~J., {Koo}, D.~C., {Martin}, C.~L., {Prochaska},
  J.~X., {Coil}, A.~L., \& {Newman}, J.~A. 2010, \apj, 719, 1503

\bibitem[{{Rupke} {et~al.}(2005){Rupke}, {Veilleux}, \& {Sanders}}]{rupke_05}
{Rupke}, D.~S., {Veilleux}, S., \& {Sanders}, D.~B. 2005, \apjs, 160, 115

\bibitem[{{Salpeter}(1955)}]{salpeter_55}
{Salpeter}, E.~E. 1955, \apj, 121, 161

\bibitem[Schaye et al.(2015)]{schaye_15} Schaye, J., Crain, 
R.~A., Bower, R.~G., et al.\ 2015, \mnras, 446, 521

\bibitem[{{Schreiber} {et~al.}(2004){Schreiber}, {Thatte}, {Eisenhauer},
  {Tecza}, {Abuter}, \& {Horrobin}}]{SchreiberJ_04a}
{Schreiber}, J., {Thatte}, N., {Eisenhauer}, F., {Tecza}, M., {Abuter}, R., \&
  {Horrobin}, M. 2004, in ASP Conf. Ser. 314: Astronomical Data Analysis
  Software and Systems (ADASS) XIII, ed. F.~{Ochsenbein}, M.~G. {Allen}, \&
  D.~{Egret}, p.380

\bibitem[{{Shen} {et~al.}(2012){Shen}, {Madau}, {Aguirre}, {Guedes}, {Mayer},
  \& {Wadsley}}]{shen_12}
{Shen}, S., {Madau}, P., {Aguirre}, A., {Guedes}, J., {Mayer}, L., \&
  {Wadsley}, J. 2012, \apj, 760, 50

\bibitem[{{Shen} {et~al.}(2013){Shen}, {Madau}, {Guedes}, {Mayer}, {Prochaska},
  \& {Wadsley}}]{shen_13}
{Shen}, S., {Madau}, P., {Guedes}, J., {Mayer}, L., {Prochaska}, J.~X., \&
  {Wadsley}, J. 2013, \apj, 765, 89

\bibitem[{{Silk}(2008)}]{silk_08}
{Silk}, J. 2008, in Astronomical Society of the Pacific Conference Series, Vol.
  390, Pathways Through an Eclectic Universe, ed. J.~H. {Knapen}, T.~J.
  {Mahoney}, \& A.~{Vazdekis}, 339
  
\bibitem[Springel et al.(2006)]{springel_06} Springel, V., Frenk, 
C.~S., \& White, S.~D.~M.\ 2006, \nat, 440, 1137

\bibitem[{{Starobinsky}(1982)}]{starobinsky_82}
{Starobinsky}, A.~A. 1982, Physics Letters B, 117, 175

\bibitem[{{Steidel}(1995)}]{steidel_95}
{Steidel}, C.~C. 1995, in QSO Absorption Lines, ed. G.~{Meylan}, 139

\bibitem[{{Steidel} {et~al.}(1994){Steidel}, {Dickinson}, \&
  {Persson}}]{steidel_94}
{Steidel}, C.~C., {Dickinson}, M., \& {Persson}, S.~E. 1994, \apjl, 437, L75

\bibitem[{{Steidel} {et~al.}(2010){Steidel}, {Erb}, {Shapley}, {Pettini},
  {Reddy}, {Bogosavljevi{\'c}}, {Rudie}, \& {Rakic}}]{steidel_10}
{Steidel}, C.~C., {Erb}, D.~K., {Shapley}, A.~E., {Pettini}, M., {Reddy}, N.,
  {Bogosavljevi{\'c}}, M., {Rudie}, G.~C., \& {Rakic}, O. 2010, \apj, 717, 289

\bibitem[{{Steidel} \& {Sargent}(1992)}]{steidel_92}
{Steidel}, C.~C., \& {Sargent}, W.~L.~W. 1992, \apjs, 80, 1

\bibitem[Tasca et 
al.(2014)]{tasca_14} Tasca, L.~A.~M., Le F{\`e}vre, O., L{\'o}pez-Sanjuan, C., et al.\ 2014, \aap, 565, AA10

\bibitem[{{van Dokkum}(2001)}]{vanDokkumP_01a}
{van Dokkum}, P.~G. 2001, \pasp, 113, 1420

\bibitem[{{Veilleux} {et~al.}(2005){Veilleux}, {Cecil}, \&
  {Bland-Hawthorn}}]{veilleux_05}
{Veilleux}, S., {Cecil}, G., \& {Bland-Hawthorn}, J. 2005, \araa, 43, 769

\bibitem[Vogelsberger et al.(2014)]{Vogelsberger_14a} Vogelsberger, M., 
Genel, S., Springel, V., et al.\ 2014, \nat, 509, 177

\bibitem[{{Vogelsberger} {et~al.}(2014){Vogelsberger}, {Genel}, {Sijacki},
  {Torrey}, {Springel}, \& {Hernquist}}]{vogelsberger_14}
{Vogelsberger}, M., {Genel}, S., {Sijacki}, D., {Torrey}, P., {Springel}, V.,
  \& {Hernquist}, L. 2014, \mnras, 438, 3607

\bibitem[{{Weiner} {et~al.}(2009){Weiner}, {Coil}, {Prochaska}, {Newman},
  {Cooper}, {Bundy}, {Conselice}, {Dutton}, {Faber}, {Koo}, {Lotz}, {Rieke}, \&
  {Rubin}}]{weiner_09}
{Weiner}, B.~J. {et~al.} 2009, \apj, 692, 187

\bibitem[{{White} \& {Frenk}(1991)}]{white_91}
{White}, S.~D.~M., \& {Frenk}, C.~S. 1991, \apj, 379, 52

\bibitem[{{White} \& {Rees}(1978)}]{white_78}
{White}, S.~D.~M., \& {Rees}, M.~J. 1978, \mnras, 183, 341

\bibitem[Wuyts et al.(2011)]{wuyts_11} Wuyts, S., F{\"o}rster 
Schreiber, N.~M., van der Wel, A., et al.\ 2011, \apj, 742, 96

\bibitem[{{Zahid} {et~al.}(2014){Zahid}, {Torrey}, {Vogelsberger}, {Hernquist},
  {Kewley}, \& {Dav{\'e}}}]{zahid_13b}
{Zahid}, H.~J., {Torrey}, P., {Vogelsberger}, M., {Hernquist}, L., {Kewley},
  L., \& {Dav{\'e}}, R. 2014, \apss, 349, 873

\bibitem[{{Zahid} {et~al.}(2013){Zahid}, {Yates}, {Kewley}, \&
  {Kudritzki}}]{zahid_13}
{Zahid}, H.~J., {Yates}, R.~M., {Kewley}, L.~J., \& {Kudritzki}, R.~P. 2013,
  \apj, 763, 92

\bibitem[{{Zaritsky} {et~al.}(2014){Zaritsky}, {Courtois}, {Mu{\~n}oz-Mateos},
  {Sorce}, {Erroz-Ferrer}, {Comer{\'o}n}, {Gadotti}, {Gil de Paz}, {Hinz},
  {Laurikainen}, {Kim}, {Laine}, {Men{\'e}ndez-Delmestre}, {Mizusawa}, {Regan},
  {Salo}, {Seibert}, {Sheth}, {Athanassoula}, {Bosma}, {Cisternas}, {Ho}, \&
  {Holwerda}}]{zaritsky_14}
{Zaritsky}, D. {et~al.} 2014, \aj, 147, 134

\end{thebibliography}
\end{document}